\documentclass[10pt, twocolumn, a4paper]{article}

\usepackage{hyperref}
    \hypersetup{colorlinks, linkcolor=black, citecolor=black, urlcolor=black}
\usepackage[top=2cm, bottom=2cm, left=1.5cm, right=1.5cm]{geometry}
\usepackage{fancyhdr}
\usepackage{microtype}
\usepackage{titlesec}
    \titleformat{\title}{\large\bfseries}{}{}{}
    \titleformat{\section}{\normalfont\bfseries}{\thesection}{0.5em}{}
    \titleformat{\subsection}{\normalfont\it}{\thesubsection}{0.5em}{}
    \titleformat{\subsubsection}{\normalfont\normalsize\it}{\thesubsubsection}{0.5em}{}
    \titleformat{\paragraph}[runin]{\normalfont\bfseries}{\theparagraph}{0.5em}{}
    \titleformat{\subparagraph}[runin]{\normalfont\normalsize\it}{\thesubparagraph}{0.5em}{}
\usepackage[font=small,labelfont=bf,labelsep=space]{caption}
\usepackage[shortlabels, inline]{enumitem} % customizable enumeration/itemization environments
\usepackage{natbib}

\usepackage{booktabs}

\usepackage{comment}
\usepackage{etoolbox}
    \newbool{PRECOMPILED}
    \booltrue{PRECOMPILED}
\usepackage{amssymb}
\usepackage{amsmath}
    \DeclareMathOperator{\dom}{dom}
    \DeclareMathOperator{\cl}{cl}
    \DeclareMathOperator{\conv}{conv}
    \DeclareMathOperator{\union}{union}
    \DeclareMathOperator{\find}{find}
\usepackage[standard]{ntheorem}
\usepackage{bbm} % more black-board bold
	\newcommand{\1}{\mathbbm{1}}
\providecommand{\1}{\mathds{1}}

\usepackage{tikz}
    \definecolor{mycolor1}{HTML}{009900}
    \definecolor{green}{HTML}{009900}
    \definecolor{cutcolor}{HTML}{FF0000}
    \definecolor{undecided}{HTML}{AFAFAF}
    \definecolor{joincolor}{HTML}{000000}
    \usetikzlibrary{decorations.pathmorphing}
	\tikzstyle{cut-edge}=[dotted]
    \tikzstyle{vertex}=[circle, draw, fill=white, inner sep=0pt, minimum width=1ex]
    \tikzset{every picture/.append style={baseline,scale=1.1}}
\usepackage{pgfplots}
\usepackage{pgfplotstable}
       
\newcommand{\imagetop}[1]{\vtop{\null\hbox{#1}}}

\newcommand{\paths}{\textnormal{-paths}}

\newcommand{\cuts}{\textnormal{-cuts}}
\newenvironment{subtikzpicture}[1]{    
    \begin{tikzpicture}[
        baseline={([yshift=-\heightof{#1}]current bounding box.north)}
    ]
}{
    \end{tikzpicture}
}

\def\clap#1{\hbox to 0pt{\hss#1\hss}}

\def\mathclap{\mathpalette\mathclapinternal}

\def\mathclapinternal#1#2{%
\clap{$\mathsurround=0pt#1{#2}$}}

\begin{document}

\title{\bf\Large Analysis \& Optimization of Graph Decompositions by Lifted Multicuts}
\author{Andrea Hor\v{n}\'akov\'a$^1$, Jan-Hendrik Lange$^1$ and Bjoern Andres$^2$}
\date{Max Planck Institute for Informatics $\cdot$ Saarland Informatics Campus $\cdot$ Saarbr\"ucken, Germany\\}

\twocolumn[
\begin{@twocolumnfalse}
\maketitle
%\vspace{-4ex} % micro-typesetting
\begin{abstract}
We study the set of all decompositions (clusterings) of a graph through its characterization as a set of lifted multicuts.
This leads us to practically relevant insights related to the definition of a class of decompositions by must-join and must-cut constraints and related to the comparison of clusterings by metrics.
To find optimal decompositions defined by minimum cost lifted multicuts, we establish some properties of some facets of lifted multicut polytopes, define efficient separation procedures and apply these in a branch-and-cut algorithm.
\end{abstract}
\vspace{5ex}
\end{@twocolumnfalse}
]

\footnotetext[1]{Authors contributed equally.}
\footnotetext[2]{Correspondence: andres@mpi-inf.mpg.de}

\section{Introduction}
    This article is about the set of all decompositions (clusterings) of a graph.
A decomposition of a graph $G = (V,E)$ is a partition $\Pi$ of the node set $V$ such that, for every subset $U \in \Pi$ of nodes, the subgraph of $G$ induced by $U$ is connected.
An example is depicted in Fig.~\ref{figure:graph-decomposition}.
Decompositions of a graph arise in practice, as feasible solutions of clustering problems, and in theory, as a generalization of partitions of a set, to which they specialize for complete graphs.

We study the set of all decompositions of a graph through its characterization as a set of multicuts.
A multicut of $G$ is a subset $M \subseteq E$ of edges such that, for every (chordless) cycle $C \subseteq E$ of $G$, we have $|M \cap C| \neq 1$.
An example is depicted in Fig.~\ref{figure:graph-decomposition}.
For any graph $G$, a one-to-one relation exists between the decompositions and the multicuts of $G$.
The multicut induced by a decomposition is the set of edges that straddle distinct components.
Multicuts are useful in the study of decompositions as the characteristic function $x \in \{0, 1\}^E$ of a multicut $x^{-1}(1)$ of $G$ makes explicit, for every pair $\{v,w\} \in E$ of neighboring nodes, whether $v$ and $w$ are in distinct components.
To make explicit also for non-neighboring nodes, specifically, for all $\{v,w\} \in E'$ with $E \subseteq E' \subseteq {V \choose 2}$, whether $v$ and $w$ are in distinct components, we define a lifting of the multicuts of $G$ to multicuts of $G' = (V, E')$.
The multicuts of $G'$ lifted from $G$ are still in one-to-one relation with the decompositions of $G$.
Yet, they are a more expressive model of these decompositions than the multicuts of $G$.
We apply lifted multicuts in three ways:

\begin{figure}
\centering
\ifbool{PRECOMPILED}{
\includegraphics[scale=1]{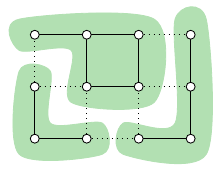}
}{
\begin{tikzpicture}[scale=0.8]
    % component 0
    \draw[draw=green!30, fill=green!30] plot[smooth cycle, tension=0.5] coordinates
        {(-0.3, 2.3) (2.3, 2.3) (2.3, 0.7) (0.7, 0.7) (0.7, 1.7) (-0.3, 1.7)};
    % component 1
    \draw[draw=green!30, fill=green!30] plot[smooth cycle, tension=0.5] coordinates
        {(-0.3, -0.3) (-0.3, 1.3) (0.3, 1.3) (0.3, 0.3) (1.3, 0.3) (1.3, -0.3)};
    % component 2
    \draw[draw=green!30, fill=green!30] plot[smooth cycle, tension=0.5] coordinates
        {(1.7, -0.3) (1.7, 0.3) (2.7, 0.3) (2.7, 2.3) (3.3, 2.3) (3.3, -0.3)};
	% edges (not cut)
	\draw (0, 0) -- (0, 1);
	\draw (0, 0) -- (1, 0);
	\draw (0, 2) -- (1, 2);
	\draw (1, 1) -- (1, 2);
	\draw (1, 1) -- (2, 1);
	\draw (1, 2) -- (2, 2);
	\draw (2, 1) -- (2, 2);
	\draw (2, 0) -- (3, 0);
	\draw (3, 0) -- (3, 1);
	\draw (3, 1) -- (3, 2);
	% cut
	\draw[style=cut-edge] (0, 1) -- (0, 2);
	\draw[style=cut-edge] (0, 1) -- (1, 1);
	\draw[style=cut-edge] (1, 0) -- (1, 1);
	\draw[style=cut-edge] (1, 0) -- (2, 0);
	\draw[style=cut-edge] (2, 0) -- (2, 1);
	\draw[style=cut-edge] (2, 1) -- (3, 1);
	\draw[style=cut-edge] (2, 2) -- (3, 2);
	% nodes, component 0
	\node[style=vertex] at (0, 0) {};
	\node[style=vertex] at (1, 0) {};
	\node[style=vertex] at (0, 1) {};
	% nodes, component 1
	\node[style=vertex] at (0, 2) {};
	\node[style=vertex] at (1, 1) {};
	\node[style=vertex] at (1, 2) {};
	\node[style=vertex] at (2, 1) {};
	\node[style=vertex] at (2, 2) {};
	% nodes, component 2
	\node[style=vertex] at (2, 0) {};
	\node[style=vertex] at (3, 0) {};
	\node[style=vertex] at (3, 1) {};
	\node[style=vertex] at (3, 2) {};
\end{tikzpicture}
}
\caption{A decomposition of a graph is a partition of the node set into connected subsets.
Above, one decomposition is depicted in green.
Any decomposition is characterized by the set of those edges (depicted as dotted lines) that straddle distinct components.
Such edge sets are precisely the multicuts of the graph.}
\label{figure:graph-decomposition}
\end{figure}
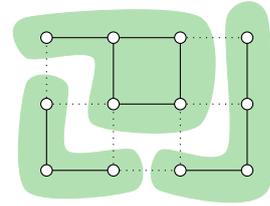

Firstly, we study problems related to the definition of a class of decompositions by \emph{must-cut} or \emph{must-join} constraints (Section~\ref{section:partial}).
The first problem is to decide whether a set of such constraints is \emph{consistent}, i.e., whether a decomposition of the given graph exists that satisfies the constraints.
We show that this decision problem is \textsc{np}-complete in general and can be solved efficienty for a subclass of constraints.
Must-cut and must-join constraints have applications where defining a decomposition totally is an ambiguous and tedious task, e.g., in image segmentation.
They relax this task to one of defining a decomposition partially.
The second problem is to decide whether a consistent set of must-join and must-cut constraints is \emph{maximally specific}, i.e., whether no such constraint can be added without changing the set of decompositions that satisfy the constraints.
We show that this decision problem is \textsc{np}-hard in general and can be solved efficienty for a subclass of constraints.
This finding is relevant for comparing the classes of decompositions definable by must-join and must-cut constraints by certain metrics, which is the next topic.

As a second application of lifted multicuts, we study the comparison of decompositions and classes of decompositions by \emph{metrics} (Section~\ref{section:metric}).
To obtain a metric on the set of all decompositions of a given graph, we define a metric on a set of lifted multicuts that characterize these decompositions.
By lifting to different graphs, we obtain different metrics, two of which are well-known and here generalized.
% This generalization has applications in image segmentation where it points to the scale(s) at which most discrepancies between two decompositions of the same image occur.
%
To extend this metric to the classes of decompositions definable by must-join and must-cut constraints, we define a metric on partial lifted multicuts that characterize these classes, connecting results of Sections~\ref{section:partial} and \ref{section:metric}.
We show that computing this metric is \textsc{np}-hard in general and efficient for a subclass of must-join and must-cut constraints.
These findings have implications on the applicability of must-join and must-cut constraints as a form of supervision, more specifically, on the practicality of certain error metrics and loss functions.

\begin{figure}[t!]
\ifbool{PRECOMPILED}{
\begin{tabular}{@{}l@{}}
\includegraphics[scale=1]{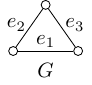}
\end{tabular}
}{
\begin{tikzpicture}
    \node[style=vertex] at (0, 0) (a) {};
    \node[style=vertex] at (1, 0) (b) {};
    \node[style=vertex] at (0.5, 0.71) (c) {};
    \draw (a) -- (b);
    \draw (b) -- (c);
    \draw (c) -- (a);
    \node at (0.5, 0.15) {$e_1$};
    \node at (0.05, 0.4) {$e_2$};
    \node at (0.95, 0.4) {$e_3$};
    \node at (0.5, -0.3) {$G$};
\end{tikzpicture}
}
\hfill
\begin{tabular}{@{}l@{\ }l@{}}
\ifbool{PRECOMPILED}{
\includegraphics[scale=1]{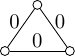}
& \includegraphics[scale=1]{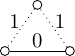} \\
\includegraphics[scale=1]{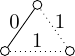}
& \includegraphics[scale=1]{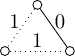} \\
\includegraphics[scale=1]{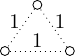}
}{
    \begin{tikzpicture}
        \node[style=vertex] at (0, 0) (a) {};
        \node[style=vertex] at (1, 0) (b) {};
        \node[style=vertex] at (0.5, 0.71) (c) {};
        \draw (a) -- (b);
        \draw (b) -- (c);
        \draw (c) -- (a);
        \node at (0.5, 0.15) {0};
        \node at (0.15, 0.45) {0};
        \node at (0.85, 0.45) {0};
    \end{tikzpicture}
&     \begin{tikzpicture}
        \node[style=vertex] at (0, 0) (a) {};
        \node[style=vertex] at (1, 0) (b) {};
        \node[style=vertex] at (0.5, 0.71) (c) {};
        \draw (a) -- (b);
        \draw[style=cut-edge] (b) -- (c);
        \draw[style=cut-edge] (c) -- (a);
        \node at (0.5, 0.15) {0};
        \node at (0.15, 0.45) {1};
        \node at (0.85, 0.45) {1};
    \end{tikzpicture} \\
    \begin{tikzpicture}
        \node[style=vertex] at (0, 0) (a) {};
        \node[style=vertex] at (1, 0) (b) {};
        \node[style=vertex] at (0.5, 0.71) (c) {};
        \draw[style=cut-edge] (a) -- (b);
        \draw[style=cut-edge] (b) -- (c);
        \draw (c) -- (a);
        \node at (0.5, 0.15) {1};
        \node at (0.15, 0.45) {0};
        \node at (0.85, 0.45) {1};
    \end{tikzpicture}
&      \begin{tikzpicture}
        \node[style=vertex] at (0, 0) (a) {};
        \node[style=vertex] at (1, 0) (b) {};
        \node[style=vertex] at (0.5, 0.71) (c) {};
        \draw[style=cut-edge] (a) -- (b);
        \draw (b) -- (c);
        \draw[style=cut-edge] (c) -- (a);
        \node at (0.5, 0.15) {1};
        \node at (0.15, 0.45) {1};
        \node at (0.85, 0.45) {0};
    \end{tikzpicture} \\
    \begin{tikzpicture}
        \node[style=vertex] at (0, 0) (a) {};
        \node[style=vertex] at (1, 0) (b) {};
        \node[style=vertex] at (0.5, 0.71) (c) {};
        \draw[style=cut-edge] (a) -- (b);
        \draw[style=cut-edge] (b) -- (c);
        \draw[style=cut-edge] (c) -- (a);
        \node at (0.5, 0.15) {1};
        \node at (0.15, 0.45) {1};
        \node at (0.85, 0.45) {1};
    \end{tikzpicture}
}
\end{tabular}
\hfill
\ifbool{PRECOMPILED}{
\begin{tabular}{@{}l@{}}
\includegraphics[scale=1]{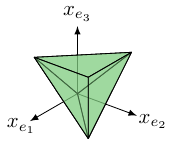}
\end{tabular}
\\[-2ex] % micro-typesetting
}{
% Sketch output, version 0.3 (build 7d, Mon Mar 5 15:32:20 2012)
% Output language: PGF/TikZ,LaTeX
\begin{tikzpicture}[line join=round]
\draw[arrows=-latex](0,0)--(.914,-.338);
\filldraw[fill=mycolor1!50,fill opacity=0.5](0,0)--(.166,-.691)--(.831,.637)--cycle;
\draw[arrows=-latex](0,0)--(0,1.039);
\filldraw[fill=mycolor1!50,fill opacity=0.5](0,0)--(-.665,.56)--(.831,.637)--cycle;
\draw[arrows=-latex](0,0)--(-.731,-.422);
\filldraw[fill=mycolor1!50,fill opacity=0.5](0,0)--(.166,-.691)--(-.665,.56)--cycle;
\filldraw[fill=mycolor1!50,fill opacity=0.5](.166,.253)--(-.665,.56)--(.831,.637)--cycle;
\filldraw[fill=mycolor1!50,fill opacity=0.5](.166,.253)--(.166,-.691)--(.831,.637)--cycle;
\filldraw[fill=mycolor1!50,fill opacity=0.5](.166,.253)--(.166,-.691)--(-.665,.56)--cycle;
 \node at (-.864,-.499) {$x_{e_1}$};  \node at (1.164,-.43) {$x_{e_2}$};  \node at (0,1.227) {$x_{e_3}$}; \end{tikzpicture}% End sketch output
\\[-1ex] % micro-typesetting
}
\caption{For any connected graph $G$ (left), 
the characteristic functions of all multicuts of $G$ (middle)
span, as their convex hull in $\mathbb{R}^E$, the \emph{multicut polytope} of $G$ (right),
a 01-polytope that is $|E|$-dimensional
\citep{chopra-1993}.}
\label{figure:multicut-polytope}
\end{figure}
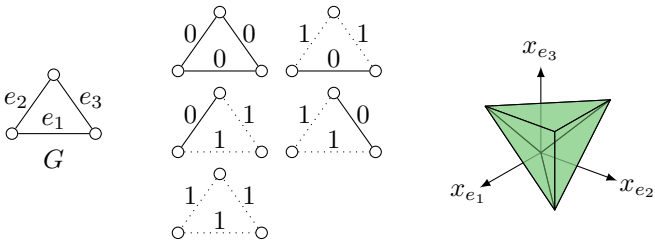

As a third application of lifted multicuts, we study the optimization of graph decompositions by minimum cost lifted multicuts.
The minimum cost lifted multicut problem is a generalization of the correlation clustering problem with applications in the field of computer vision.
To tackle this problem, we establish some properties of some facets of lifted multicut polytopes (Fig.~\ref{figure:multicut-polytope} and \ref{figure:lifted-multicut-polytope}), define efficient separation procedures and apply these in a branch-and-cut algorithm.

\subsection{Related Work}
    % !TeX root = 0000.tex

Initial motivation to study decompositions of a graph by multicuts came from the field of polyhedral optimization.
Multicut polytopes are studied by 
\citet{groetschel-1989,deza-1991,deza-1992,chopra-1993,chopra-1995} and \citet{deza-1997}
who characterize several classes of their facets.

The binary linear program whose feasible solutions are all multicuts of a graph is known as the correlation clustering problem from the work of \citet{bansal-2004} and \citet{demaine-2006} who establish its \textsc{apx}-hardness and a logarithmic approximation.
The stability of its solutions is analyzed by \citet{nowozin-2009}.
Generalizations to polynomial objective functions are studied by \citet{kim-2014} and \citet{kappes-2016}.
Interestingly, the problem remains \textsc{np}-hard for planar graphs \citep{voice-2012,bachrach-2013} where it admits a PTAS \citep{klein-2015} and relaxations that are often tight in practice \citep{yarkony-2012}.

The lifting of multicuts we define makes path connectedness explicit.
For a single component, this is studied by \citet{nowozin-2010} who introduce the connected subgraph polytope and outer relaxations.

Applications of the minimum cost lifted multicut problem and experimental comparisons to the correlation clustering problem in the field of computer vision are by \cite{keuper-2015a} and \cite{tang-2017} who find feasible solutions by local search \citep{keuper-2015a,levinkov-2017}, and by \cite{beier-2017} who find feasible solutions by consensus optimization \citep{beier-2016}.

The complexity of several decision problems related to clustering with must-join and must-cut constraints is established by \citet{davidson-2007}.

Well-known metrics on the set of all decompositions of a graph are the metric of \citet{rand-1971} and the variation of information \citep{meila-2007}.
\section{Multicuts}
    % !TeX root = 0000.tex

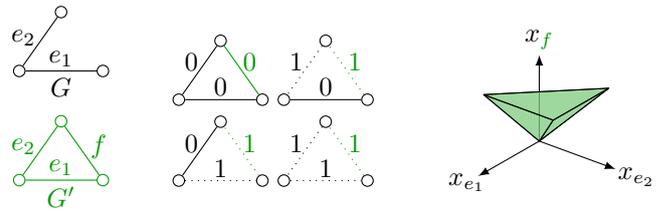
\begin{figure}[t!]
\begin{tabular}{@{}l@{}}
\ifbool{PRECOMPILED}{
\includegraphics[scale=1]{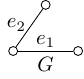}
\\
\includegraphics[scale=1]{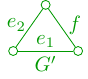}
}{
    \begin{tikzpicture}
        \node[style=vertex] at (0, 0) (a) {};
        \node[style=vertex] at (1, 0) (b) {};
        \node[style=vertex] at (0.5, 0.71) (c) {};
        \draw (a) -- (b);
        \draw (c) -- (a);
        \node at (0.5, 0.15) {$e_1$};
        \node at (0.05, 0.4) {$e_2$};
        \node at (0.5, -0.2) {$G$};
    \end{tikzpicture}
\\[3ex]
    \begin{tikzpicture}
        \color{mycolor1}        
        \node[style=vertex] at (0, 0) (a) {};
        \node[style=vertex] at (1, 0) (b) {};
        \node[style=vertex] at (0.5, 0.71) (c) {};
        \draw (a) -- (b);
        \draw[mycolor1] (b) -- (c);
        \draw (c) -- (a);
        \node at (0.5, 0.15) {$e_1$};
        \node at (0.05, 0.4) {$e_2$};
        \node at (0.95, 0.4) {\color{mycolor1}$f$};
        \node at (0.5, -0.2) {$G'$};
    \end{tikzpicture}
}
\end{tabular}
\hfill
\begin{tabular}{@{}l@{\ }l@{}}
\ifbool{PRECOMPILED}{
\includegraphics[scale=1]{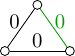}
& \includegraphics[scale=1]{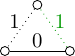} \\
\includegraphics[scale=1]{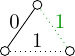}
& \includegraphics[scale=1]{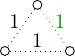}
}{
    \begin{tikzpicture}
        \node[style=vertex] at (0, 0) (a) {};
        \node[style=vertex] at (1, 0) (b) {};
        \node[style=vertex] at (0.5, 0.71) (c) {};
        \draw (a) -- (b);
        \draw[mycolor1] (b) -- (c);
        \draw (c) -- (a);
        \node at (0.5, 0.15) {0};
        \node at (0.15, 0.45) {0};
        \node at (0.85, 0.45) {\color{mycolor1}0};
    \end{tikzpicture}
&      \begin{tikzpicture}
        \node[style=vertex] at (0, 0) (a) {};
        \node[style=vertex] at (1, 0) (b) {};
        \node[style=vertex] at (0.5, 0.71) (c) {};
        \draw (a) -- (b);
        \draw[style=cut-edge, mycolor1] (b) -- (c);
        \draw[style=cut-edge] (c) -- (a);
        \node at (0.5, 0.15) {0};
        \node at (0.15, 0.45) {1};
        \node at (0.85, 0.45) {\color{mycolor1}1};
    \end{tikzpicture} \\
    \begin{tikzpicture}
        \node[style=vertex] at (0, 0) (a) {};
        \node[style=vertex] at (1, 0) (b) {};
        \node[style=vertex] at (0.5, 0.71) (c) {};
        \draw[style=cut-edge] (a) -- (b);
        \draw[style=cut-edge, mycolor1] (b) -- (c);
        \draw (c) -- (a);
        \node at (0.5, 0.15) {1};
        \node at (0.15, 0.45) {0};
        \node at (0.85, 0.45) {\color{mycolor1}1};
    \end{tikzpicture}
&  \begin{tikzpicture}
        \node[style=vertex] at (0, 0) (a) {};
        \node[style=vertex] at (1, 0) (b) {};
        \node[style=vertex] at (0.5, 0.71) (c) {};
        \draw[style=cut-edge] (a) -- (b);
        \draw[style=cut-edge, mycolor1] (b) -- (c);
        \draw[style=cut-edge] (c) -- (a);
        \node at (0.5, 0.15) {1};
        \node at (0.15, 0.45) {1};
        \node at (0.85, 0.45) {\color{mycolor1}1};
    \end{tikzpicture}
}
\end{tabular}
\hfill
\ifbool{PRECOMPILED}{
\begin{tabular}{@{}l@{}}
\includegraphics[scale=1]{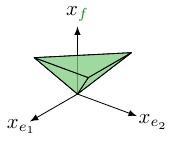}
\end{tabular}
}{
% Sketch output, version 0.3 (build 7d, Mon Mar 5 15:32:20 2012)
% Output language: PGF/TikZ,LaTeX
\begin{tikzpicture}[line join=round]
\draw[arrows=-latex](0,0)--(-.731,-.422);
\draw[arrows=-latex](0,0)--(.914,-.338);
\draw[arrows=-latex](0,0)--(0,1.039);
\filldraw[fill=mycolor1!50,fill opacity=0.5](0,0)--(.831,.637)--(-.665,.56)--cycle;
\filldraw[fill=mycolor1!50,fill opacity=0.5](0,0)--(.831,.637)--(.166,.253)--cycle;
\filldraw[fill=mycolor1!50,fill opacity=0.5](0,0)--(-.665,.56)--(.166,.253)--cycle;
\filldraw[fill=mycolor1!50,fill opacity=0.5](.831,.637)--(-.665,.56)--(.166,.253)--cycle;
 \node at (-.864,-.499) {$x_{e_1}$};  \node at (1.164,-.43) {$x_{e_2}$};  \node at (0,1.227) {$x_{\color{mycolor1}f}$}; \end{tikzpicture}% End sketch output
}
\caption{For any connected graph $G = (V, E)$ (top left)
and any graph $G' = (V, E')$ with $E \subseteq E'$ (bottom left),
those multicuts of $G'$ that are lifted from $G$ (middle)
span, as their convex hull in $\mathbb{R}^E$, the \emph{lifted multicut polytope} w.r.t.~$G$ and $G'$ (right),
a 01-polytope that is $|E'|$-dimensional
(Thm.~\ref{theorem:dimension}).}
\label{figure:lifted-multicut-polytope}
\end{figure}

\begin{definition}
\label{definition:decomposition}
Let $G = (V,E)$ be any graph.
A subgraph $G' = (V',E')$ of $G$
is called a \emph{component} of $G$ 
iff $G'$ is non-empty, node-induced\footnote{That is: $E' = E \cap {V' \choose 2}$}, and connected\footnote{We do not require a component to be maximal w.r.t.~the subgraph relation.}.
A partition $\Pi$ of $V$ is called a \emph{decomposition} of $G$
iff, for every $U \in \Pi$, the subgraph $(U, E \cap {U \choose 2})$ of $G$ induced by $U$ is connected (and hence a component of $G$).
\end{definition}

For any graph $G$, we denote by $D_G \subset 2^{2^V}$ the set of all decompositions of $G$.
Useful in the study of decompositions are the multicuts of a graph:

\begin{definition}
\label{definition:multicut}
For any graph $G = (V,E)$, a subset $M \subseteq E$ of edges is called a \emph{multicut} of $G$ iff, for every cycle $C \subseteq E$ of $G$, we have $|C \cap M| \neq 1$.
\end{definition}

\begin{lemma}\citep{chopra-1993}
It is sufficient in 
Def.~\ref{definition:multicut} 
to consider only the \emph{chordless} cycles.
\end{lemma}

For any graph $G$, we denote by $M_G \subseteq 2^E$ the set of all multicuts of $G$.
One reason why multicuts are useful in the study of decompositions is that, 
for every graph $G$, a one-to-one relation exists between the decompositions and the multicuts of $G$.
An example is depicted in Fig.~\ref{figure:graph-decomposition}:

\begin{lemma}
\label{lemma:characteristic}
For any graph $G = (V, E)$, the map $\phi_G: D_G \to 2^E$
defined by \eqref{eq:map-phi} is a bijection from $D_G$ to $M_G$.
\begin{align} 
& \forall \Pi \in D_G\ 
    \forall \{v,w\} \in E:\nonumber\\
& \quad
    \{v,w\} \in \phi_G(\Pi) 
    \ \Leftrightarrow \ 
    \forall U \in \Pi (v \notin U \vee w \notin U)
\label{eq:map-phi}
\end{align}
\end{lemma}

Another reason why multicuts are useful in the study of decompositions 
is that, for any graph $G = (V,E)$ and any decomposition $\Pi$ of $G$, 
the characteristic function of the multicut induced by $\Pi$ is a 01-encoding of $\Pi$ of fixed length $|E|$.

\begin{lemma}\citep{chopra-1993}
\label{lemma:encoding}
For any graph $G = (V, E)$ and any $x \in \{0,1\}^E$, the set $x^{-1}(1)$ of those edges that are labeled 1 is a multicut of $G$ iff \eqref{eq:cycle-inequalities} holds.
It is sufficient in \eqref{eq:cycle-inequalities} to consider only \emph{chordless} cycles.
\begin{align}
\forall C \in \textnormal{cycles}(G)\ 
    \forall e \in C:\ 
        x_e \leq \hspace{-2ex} \sum_{e' \in C \setminus \{e\}} \hspace{-2ex} x_{e'}    
\label{eq:cycle-inequalities}
\end{align}
\end{lemma}

For any graph $G = (V, E)$, we denote by $X_G$ the set of all $x \in \{0,1\}^E$ that satisfy 
\eqref{eq:cycle-inequalities}.

\subsection{Complete Graphs}

The decompositions of a complete graph $K_V := (V, {V \choose 2})$ are precisely the partitions of the node set $V$
(by Def.~\ref{definition:decomposition}).
The multicuts of a complete graph $K_V$ relate one-to-one to the equivalence relations on $V$:

\begin{lemma}
\label{lemma:multicut-complete-graph}
For any set $V$ and the complete graph $K_V$,
the map $\psi: M_{K_V} \to 2^{V \times V}$ 
defined by \eqref{eq:map-to-eqr}
is a bijection between $M_{K_V}$ and the set of all equivalence relations on $V$.
\begin{align}
& \forall M \in M_{K_V}\ 
    \forall v, w \in V:\nonumber\\
& \quad 
    (v, w) \in \psi(M) 
    \ \Leftrightarrow \ 
    \{v, w\} \notin M
    \label{eq:map-to-eqr}
\end{align}
\end{lemma}

The bijection between the decompositions of a graph and the multicuts of a graph
(Lemma~\ref{lemma:characteristic})
specializes, for complete graphs, to the well-known bijection between the partitions of a set and the equivalence relations on the set
(by Lemma~\ref{lemma:multicut-complete-graph}).
In this sense, decompositions and multicuts of graphs generalize partitions of sets and equivalence relations on sets.

\section{Lifted Multicuts}
    \label{section:lifting}
    % !TeX root = 0000.tex

For any graph $G = (V,E)$, 
the characteristic function $x \in X_G$ of a multicut $x^{-1}(1)$ of $G$ makes explicit, 
for every pair $\{v,w\} \in E$ of neighboring nodes,
whether $v$ and $w$ are in distinct components.
To make explicit also for non-neighboring nodes, 
specifically, for all $\{v,w\} \in E'$ with $E \subseteq E' \subseteq \tbinom{V}{2}$, 
whether $v$ and $w$ are in distinct components, 
we define a lifting of the multicuts of $G$ to multicuts of $G' = (V, E')$:

\begin{definition}
\label{definition:lifting}
For any graphs $G = (V, E)$ and $G' = (V, E')$ with $E \subseteq E'$,
the composed map $\lambda_{GG'} := \phi_{G'} \circ \phi_G^{-1}$
is called the \emph{lifting} of multicuts from $G$ to $G'$.
\end{definition}

For any graphs $G = (V, E)$ and $G' = (V, E')$ with $E \subseteq E'$, we introduce the notation $F_{GG'} := E' \setminus E$, for brevity.

\begin{lemma}
\label{lemma:lifted-multicut}
For any connected graph $G = (V, E)$, 
any graph $G' = (V, E')$ with $E \subseteq E'$
and any $x \in \{0,1\}^{E'}$, 
the set $x^{-1}(1)$ is a multicut of $G'$ lifted from $G$ iff
\begin{align}
& \forall C \in \textnormal{cycles}(G) \, \forall e \in C: \, 
    x_e \leq \hspace{-2ex} \sum_{e' \in C \setminus \{e\}} \hspace{-2ex} x_{e'} 
    \label{eq:lmc-cycle}\\
& \forall vw \in F_{GG'} \, \forall P \in vw\paths(G): \, 
   	x_{vw} \leq \sum_{e \in P} x_e
    \label{eq:lmc-path}\\
& \forall vw \in F_{GG'} \, \forall C \in vw\cuts(G): \, 
	1 - x_{vw} \leq \sum_{e \in C} (1 - x_e) 
    \label{eq:lmc-cut}
\end{align}
\end{lemma}

For any graphs $G = (V, E)$ and $G' = (V, E')$ with $E \subseteq E'$
we denote by $X_{GG'}$ the set of all $x \in \{0,1\}^{E'}$ that satisfy
\eqref{eq:lmc-cycle}--\eqref{eq:lmc-cut}.

\section{Partial Lifted Multicuts} \label{section:partial}
    % !TeX root = 0000.tex

As a first application of lifted multicuts, we study the class of decompositions of a graph definable by must-join and must-cut constraints.
For this, we consider partial functions.
For any set $E$, a partial characteristic function of subsets of $E$ is a function from any subset $F \subseteq E$ to $\{0,1\}$.
With some abuse of notation, we denote the set of all partial characteristic functions of subsets of $E$ by $\{0,1,*\}^E := \bigcup_{F \subseteq E} \{0,1\}^F$.
For any $x \in \{0,1,*\}^E$, we denote the domain of $x$ by $\dom x := x^{-1}(\{0,1\})$.

For any connected graph $G = (V, E)$ whose decompositions we care about and any graph $G' = (V, E')$ with $E \subseteq E'$, we consider a partial function $\tilde x \in \{0,1,*\}^{E'}$.
For any $\{v,w\} \in \dom \tilde x$, we constrain the nodes $v$ and $w$ to the same component if $\tilde x_{vw} = 0$ and to distinct components if $\tilde x_{vw} = 1$.

\subsection{Consistency}\label{section:consistency}

A natural question to ask is whether a decomposition of the graph $G$ exists that satisfies these constraints.
We show that this decision problem is \textsc{np}-complete.

\begin{definition}
For any connected graph $G = (V, E)$,
any graph $G' = (V, E')$ with $E \subseteq E'$,
and any $\tilde x \in \{0,1,*\}^{E'}$,
the elements of
\begin{align}
X_{GG'}[\tilde x] := \left\{ 
    x \in X_{GG'} \,\middle|\, \forall e \in \dom \tilde x: \, x_e = \tilde x_e 
\right\}
\end{align}
are called the \emph{completions} of $\tilde x$ in $X_{GG'}$.
In addition, $\tilde x$ is called \emph{consistent} and a \emph{partial characterization of multicuts of $G'$ lifted from $G$} iff
\begin{align}
X_{GG'}[\tilde x] \neq \emptyset
\enspace .
\label{eq:consistency}
\end{align}
\end{definition}

We denote the set of all partial characterizations of multicuts of $G'$ lifted from $G$ by
\begin{align}
\tilde X_{GG'} & := \left\{
    \tilde x \in \{0,1,*\}^{E'} \,\middle|\, X_{GG'}[\tilde x] \neq \emptyset 
\right\}
\enspace .
\end{align}

\begin{theorem}\label{theorem:consistency}
Deciding consistency is \textsc{np}-complete.
\end{theorem}

\begin{lemma}\label{lemma:polynomialConsistency}
Consistency can be decided efficiently if $E \subseteq \dom \tilde x$ or 
\begin{align}
& \forall vw \in \dom \tilde x \setminus E: \nonumber \\
& \quad \tilde x_{vw} = 1 
\vee \exists P \in vw\textnormal{-path}(G) \,
    \forall e \in P: \,
        \tilde x_e = 0 
\label{eq:supportedPaths}
\end{align}
\end{lemma}

\subsection{Specificity}\label{section:specificity}

A less obvious question to ask for any partial characterization $\tilde x$ of multicuts of $G'$ lifted from $G$ is whether $\tilde x$ is maximally specific for its completions in $X_{GG'}$.
In other words, is there no edge $e \in E' \setminus \dom \tilde x$ such that, for any completions $x,x'$ of $\tilde x$ in $X_{GG'}$, we have $x_e = x'_e$, i.e., an edge that could be included in $\dom \tilde x$ without changing the set of completions of $\tilde x$ in $X_{GG'}$?
We show that deciding maximal specificity is \textsc{np}-hard.

\begin{definition}
\label{definition:specificity}
Let $G = (V, E)$ a connected graph and $G' = (V, E')$ a graph with $E \subseteq E'$.
For any $\tilde x \in \tilde X_{GG'}$, the edges
\begin{align}
E'[\tilde x] := \left\{
    e \in E'
    \,\middle|\,
    \forall x,x' \in X_{GG'}[\tilde x]: \,
        x_e = x'_e
\right\}
\end{align}
are called \emph{decided}. 
The edges $E' \setminus E'[\tilde x]$ are called \emph{undecided}.
Moreover, $\tilde x$ is called \emph{maximally specific} iff\footnote{Note that \eqref{eq:closedness} is equivalent to $E'[\tilde x] = \dom \tilde x$, as $E'[\tilde x] \supseteq \dom \tilde x$ holds by definition of $E'[\tilde x]$.}
\begin{align}
E'[\tilde x] \subseteq \dom \tilde x
\enspace .
\label{eq:closedness}
\end{align}
\end{definition}

\begin{theorem}\label{theorem:specificity}
Deciding maximal specificity is \textsc{np}-hard.
\end{theorem}

\begin{lemma}\label{theorem:specificityPolynomial}
Maximal specificity can be decided efficiently if $E' = E$ or $E \subseteq \dom \tilde x$.
\end{lemma}

Below, we justify the term \emph{maximal specificity} and define an operation that maps any partial characterization of lifted multicuts to one that is maximally specific.

\begin{definition}
For any connected graph $G = (V, E)$ and any graph $G' = (V, E')$ with $E \subseteq E'$,
the relation $\leq$ on $\tilde X_{GG'}$ defined by \eqref{eq:specificity} is called the \emph{specificity} of partial characterizations of multicuts of $G'$ lifted from $G$.
\begin{align}
& \forall \tilde x, \tilde x' \in \tilde X_{GG'}: \label{eq:specificity}\\
& \quad \tilde x \leq \tilde x' 
\,\Leftrightarrow\,
\dom \tilde x \subseteq \dom \tilde x'
\wedge
\forall e \in \dom \tilde x: \tilde x_e = \tilde x'_e
\nonumber
\end{align}
\end{definition}

\begin{lemma}\label{lemma:partialOrder}
For any connected graph $G = (V, E)$ and any graph $G' = (V, E')$ with $E \subseteq E'$,
specificity is a partial order on $\tilde X_{GG'}$.
\end{lemma}

Note that two partial characterizations $\tilde x, \tilde x' \in \tilde X_{GG'}$ with the same completions $X_{GG'}[\tilde x] = X_{GG'}[\tilde x']$ need not be comparable w.r.t.~$\leq$.
For example, consider the graphs $G, G'$ from Fig.~\ref{figure:lifted-multicut-polytope}, consider $\tilde x: e_1 \mapsto 0, e_2 \mapsto 0$ and $\tilde x': f \mapsto 0$.
Nevertheless, we have the following lemma.

\begin{lemma}\label{lemma:maximalSpecificity}
For any connected graph $G = (V, E)$,
any graph $G' = (V, E')$ with $E \subseteq E'$,
any $\tilde x \in \tilde X_{GG'}$ and
\begin{align}
\tilde X_{GG'}[\tilde x] :=
\left\{
    \tilde x' \in \tilde X_{GG'}
    \,\middle|\,
    X_{GG'}[\tilde x'] = X_{GG'}[\tilde x]
\right\}
\end{align}
a maximum of $\tilde X_{GG'}[\tilde{x}]$ w.r.t.~$\leq$ exists and is unique.
Moreover, $\tilde x$ is maximally specific in the sense of Def.~\ref{definition:specificity} 
iff $\tilde x$ is maximal w.r.t.~$\leq$ in $\tilde X_{GG'}[\tilde x]$.
\end{lemma}

\begin{definition}\label{definition:closure}
Let $G = (V, E)$ be a connected graph and let $G' = (V, E')$ be a graph with $E \subseteq E'$.
For any $\tilde x \in \tilde X_{GG'}$, we call the unique maximum of $\tilde X_{GG'}[\tilde x]$ w.r.t.~$\leq$ the \emph{closure} of $\tilde x$ w.r.t.~$G$ and $G'$ and denote it by $\cl_{GG'} \tilde x$.
\end{definition}

We denote by $\hat X_{GG'}$ the set of all maximally specific partial characterizations of multicuts of $G'$ lifted from $G$, i.e.:
\begin{align}
\hat X_{GG'} := \left\{
    \tilde x \in \tilde X_{GG'} 
    \,\middle|\,
    \tilde x = \cl_{GG'} \tilde x
\right\}
\enspace .
\end{align}

\begin{theorem}\label{theorem:closureEquivalence}
For any $\tilde x, \tilde x' \in \tilde X_{GG'}$, we have
$X_{GG'}[\tilde x] = X_{GG'}[\tilde x'] \Leftrightarrow \tilde X_{GG'}[\tilde x] = \tilde X_{GG'}[\tilde x'] \Leftrightarrow \cl_{GG'} \tilde x = \cl_{GG'} \tilde x'$.
\end{theorem}

\begin{lemma}\label{lemma:closureLifting}
For any connected graph $G = (V, E)$,
any graph $G' = (V, E')$ with $E \subseteq E'$ and 
any $x \in X_G$,
the closure $y := \cl_{GG'} x$ of $x$ w.r.t.~$G$ and $G'$ coincides with the lifting of the multicut $x^{-1}(1)$ of $G$ to the multicut $y^{-1}(1)$ of $G'$, i.e.
\begin{align}
(\cl_{GG'} x)^{-1}(1) = \lambda_{GG'}(x^{-1}(1))
\enspace .
\end{align}
\end{lemma}

\begin{theorem}\label{theorem:computing-closure}
Computing closures is \textsc{np}-hard.
\end{theorem}

\begin{lemma}\label{lemma:closurePolynomial}
In the special case that $E' = E$ or $E \subseteq \dom \tilde x$, the closure can be computed efficiently.
\end{lemma}

\section{Metrics} \label{section:metric}
    % !TeX root = 0000.tex

\subsection{Metrics on Decompositions}

As a second application of lifted multicuts, we compare decompositions of a given graph by comparing lifted multicuts that characterize these decompositions.
We compare these lifted multicuts by comparing their characteristic functions by Hamming metrics:
For any $E \neq \emptyset$ and any $e \in E$, we define $d^1_e, d^1_E: \{0,1\}^E \times \{0,1\}^E \to \mathbb{N}_0^+$ by the forms
\begin{align}
d^1_e(x, x') & = \begin{cases}
    0 & \textnormal{if} \, x_e = x'_e \\
    1 & \textnormal{otherwise}
\end{cases} \\
d^1_E(x, x') & = \sum_{e' \in E} d^1_{e'}(x,x') \label{eq:hamming}
\enspace .
\end{align}

\begin{theorem}\label{theorem:metric}
For any connected graph $G = (V, E)$,
any graph $G' = (V, E')$,
any $\mu: E' \to \mathbb{R}^+$,
the set $E'' := E \cup E'$ and 
the graph $G'' := (V, E'')$,
the function $d_{E'}^\mu: X_{GG''} \times X_{GG''} \to \mathbb{R}_0^+$ of the form \eqref{eq:metric-total} is a pseudo-metric on $X_{GG''}$. 
Iff $G'$ is a supergraph of $G$, i.e., iff $E \subseteq E'$, $d_{E'}^\mu$ is a metric on $X_{GG''}$.
\begin{align}
d_{E'}^\mu(x, x') := \sum_{e \in E'} \mu_e \, d^1_e(x, x') 
\label{eq:metric-total}
\end{align}
\end{theorem}

By the one-to-one relation between decompositions and multicuts (Lemma~\ref{lemma:characteristic}), $d^\mu_{E'}$ induces a pseudo-metric on the set $D_G$ of all decompositions of $G$.
Two special cases are well-known:
For $E' = E$ and $\mu = 1$, we have $d_{E'}^\mu = d_{E}^1$, which is the Hamming metric \eqref{eq:hamming} on the multicuts that characterize the decompositions, also known as the boundary metric on decompositions.
For $E' = \tbinom{V}{2}$ and $\mu = 1$, $d_{E'}^1$ specializes to the metric of \citet{rand-1971}.
Between these extremes, i.e., for $E \subseteq E' \subseteq \tbinom{V}{2}$, the metric $d_{E'}^\mu$ can be used to analyze more specifically how two decompositions of the same graph differ.
We propose an analysis w.r.t.~the distance $\delta_{vw}$ of nodes $v$ and $w$ in $G$, i.e., w.r.t.~the length of a shortest $vw$-path in $G$.
For this, we denote by $\delta_G := \max \{\delta_{vw} : vw \in \tbinom{V}{2}\}$ the diameter of $G$.

\begin{definition}
For any connected graph $G = (V, E)$ and any $n \in \mathbb{N}$, let $E[n] := \{vw \in \tbinom{V}{2} \,|\, \delta_{vw} = n \}$ the set of pairs of nodes of distance $n$ in $G$.
Moreover, let $\mu^n: E[n] \to \mathbb{Q}^+$ the constant function that maps any $vw \in E[n]$ to $1/|E[n]|$. 
For any connected graph $G = (V, E)$, we call the sequence 
\begin{align}
\left(d^{\mu^n}_{E[n]}\right)_{n \in \{1, \ldots, \delta_G\}}
\label{eq:spectrum}
\end{align}
the \emph{spectrum of pseudo-metrics} on decompositions of $G$.
For $E' := \tbinom{V}{2}$ and $\mu: E' \to \mathbb{Q}^+: vw \mapsto 1/(\delta_G |E[\delta_{vw}]|)$, we call the metric $d^\mu_{E'}$ the \emph{$\delta$-metric} on decompositions of $G$.
\end{definition}

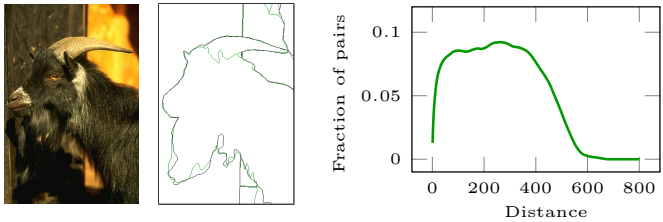
\begin{figure}
\includegraphics[width=0.2\columnwidth]{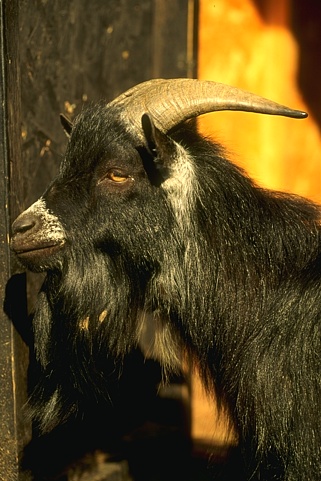}
\hfill
\frame{\includegraphics[width=0.2\columnwidth]{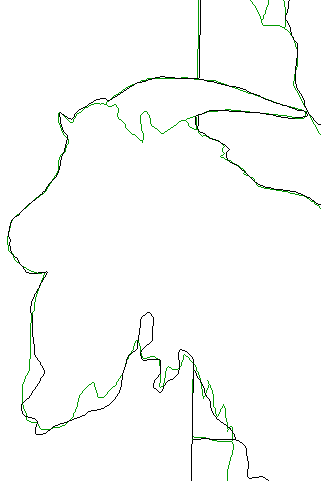}}
\hfill
\ifbool{PRECOMPILED}{
\raisebox{-2ex}{\includegraphics[scale=1.1]{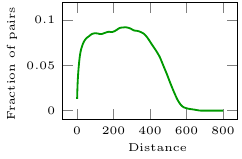}}
}{
\raisebox{2.5ex}{\begin{tikzpicture}
\begin{axis}[
    width=0.5\columnwidth,
    height=0.4\columnwidth,
    xlabel=Distance,
    xlabel style={yshift=2ex,font=\tiny},
    ylabel={Fraction of pairs},
    ylabel style={xshift=-1ex,yshift=-3.5ex,font=\tiny},
    ymin=-0.01,
    ymax=0.12,
    no markers,
    scaled ticks=false,
    tick label style={/pgf/number format/fixed},
    tick label style={font=\tiny}
]
\small
\addplot[thick, mycolor1] table[x=distance,y=spectrum]{66039-spectrum.csv};
\end{axis}
\end{tikzpicture}}
}
\caption{Depicted are two decompositions of the pixel grid graph of an image, all from \cite{arbelaez-2011}, along with the spectrum of pseudo-metrics of these decompositions.}
\label{figure:spectrum}
\end{figure}

An example of a spectrum of pseudo-metrics is depicted in Fig.~\ref{figure:spectrum}.
For any two decompositions $\Pi, \Pi'$ of a connected graph $G$ and suitable lifted multicuts $x,x'$ characterizing these decompositions, $d^{\mu^n}_{E[n]}(x, x')$ equals the fraction of pairs of nodes at distance $n$ in $G$ that are either cut by $\Pi$ and joined by $\Pi'$, or cut by $\Pi'$ and joined by $\Pi$.
I.e., the pseudo-metric $d^{\mu^n}_{E[n]}$ compares decompositions of $G$ specifically w.r.t.~the distance $n$ in $G$.
The $\delta$-metric compares decompositions w.r.t.~all distances, and each distance is weighted equally. 
This is in contrast to Rand's metric which is also a comparison w.r.t.~all distances but each distance is weighted by the number of pairs of nodes that have this distance.

\subsection{Metrics on Classes of Decompositions}

We compare classes of decompositions definable by must-join and must-cut constraints by comparing partial lifted multicuts that characterize these decompositions.
To compare partial lifted multicuts, we compare their partial characteristic functions by an extension of the Hamming metric:
For any $E \neq \emptyset$, any $e \in E$ and any $\theta \in \mathbb{R}_0^+$, we define $d^\theta_e, d^\theta_E: \{0,1,*\}^E \times \{0,1,*\}^E \to \mathbb{R}_0^+$ such that for all $\tilde x, \tilde x' \in \{0,1,*\}^E$:
\begin{align}
d^\theta_e(\tilde x, \tilde x') & = \begin{cases}
    1 & \textnormal{if } \, e \in \dom \tilde x 
    \wedge e \in \dom \tilde x' 
    \wedge \tilde x_e \neq \tilde x'_e \\
    0 & \textnormal{if } \, e \in \dom \tilde x \wedge e \in \dom \tilde x' \wedge \tilde x_e = \tilde x'_e \\   
    0 & \textnormal{if } \, e \notin \dom \tilde x \wedge e \notin \dom \tilde x' \\ 
    \theta & \textnormal{otherwise}
\end{cases}\\
d^\theta_E(\tilde x, \tilde x') & = \sum_{e' \in E} d^\theta_{e'}(\tilde x, \tilde x')
\enspace .
\end{align}

\begin{theorem}\label{theorem:metric-partial}
For any connected graph $G = (V, E)$,
any graph $G' = (V, E')$ with $E \subseteq E'$
and any $\theta \in [\tfrac{1}{2}, 1]$,
the function $\tilde d^\theta_{E'}: \tilde X_{GG'} \times \tilde X_{GG'} \to \mathbb{R}_0^+$ of the form 
\begin{align}
\tilde d^\theta_{E'}(\tilde x, \tilde x') := d^\theta_{E'}(\cl_{GG'} \tilde x, \cl_{GG'} \tilde x') 
\label{eq:metric-partial}
\end{align}
is a pseudo-metric on $\tilde X_{GG'}$ and a metric on $\hat X_{GG'}$.
Moreover, for any $\tilde x, \tilde x' \in \tilde X_{GG'}$:
\begin{align}
\tilde X_{GG'}[\tilde x] = \tilde X_{GG'}[\tilde x']
\,\Leftrightarrow\,
\tilde d^\theta_{E'}(\tilde x, \tilde x') = 0
\enspace .
\label{eq:invariance}
\end{align}
\end{theorem}

By the one-to-one relation between decompositions and multicuts (Lemma~\ref{lemma:characteristic}), 
every partial characterization of a lifted multicut $\tilde x \in \tilde X_{GG'}$ defines a class of decompositions of the graph $G$, namely those defined by the lifted multicuts characterized by $X_{GG'}[\tilde x]$.
By Theorem~\ref{theorem:metric-partial}, $\tilde d^\theta_{E'}$ with $\theta \in [\frac{1}{2}, 1]$ well-defines a metric on these classes of decompositions and hence a means of comparing the classes of decompositions definable by must-join and must-cut constraints.
Computing $\tilde d^\theta_{E'}(x,x')$ involves computing the closures of $x$ and $x'$ and is therefore \textsc{np}-hard (by Theorem~\ref{theorem:computing-closure}).

\section{Polyhedral Optimization}
    % !TeX root = 0000.tex

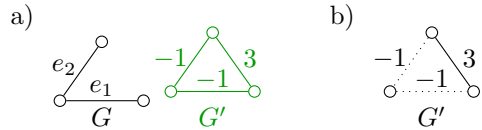
\begin{figure}
\ifbool{PRECOMPILED}{
a) \imagetop{\includegraphics[scale=1]{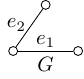}
\hspace{-2ex}
\includegraphics[scale=1]{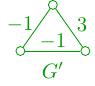}}
\hspace{4ex}
b) \imagetop{\includegraphics[scale=1]{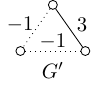}}
}{
a) \imagetop{\begin{tikzpicture}
    \node[style=vertex] at (0, 0) (a) {};
    \node[style=vertex] at (1, 0) (b) {};
    \node[style=vertex] at (0.5, 0.71) (c) {};
    \draw (a) -- (b);
    \draw (c) -- (a);
    \node at (0.5, 0.15) {$e_1$};
    \node at (0.05, 0.4) {$e_2$};
    \node at (0.5, -0.2) {$G$};
\end{tikzpicture}
\hspace{-2ex}
\begin{tikzpicture}
    \color{mycolor1}        
    \node[style=vertex] at (0, 0) (a) {};
    \node[style=vertex] at (1, 0) (b) {};
    \node[style=vertex] at (0.5, 0.71) (c) {};
    \draw (a) -- (b);
    \draw[mycolor1] (b) -- (c);
    \draw (c) -- (a);
    \node at (0.5, 0.15) {$-1$};
    \node at (0.05, 0.4) {$-1$\ \ };
    \node at (0.95, 0.4) {3};
    \node at (0.5, -0.3) {$G'$};
\end{tikzpicture}}
\hspace{4ex}
b) \imagetop{\begin{tikzpicture}
    \node[style=vertex] at (0, 0) (a) {};
    \node[style=vertex] at (1, 0) (b) {};
    \node[style=vertex] at (0.5, 0.71) (c) {};
    \draw[style=cut-edge] (a) -- (b);
    \draw (b) -- (c);
    \draw[style=cut-edge] (c) -- (a);
    \node at (0.5, 0.15) {$-1$};
    \node at (0.05, 0.4) {$-1$\ \ };
    \node at (0.95, 0.4) {$3$};
    \node at (0.5, -0.3) {$G'$};
\end{tikzpicture}}
}
\caption{Depicted above in (a) is an instance of the minimum cost lifted multicut problem (Def.~\ref{definition:lmp}) w.r.t.~graphs $G$, $G'$ and costs $c = (-1, -1, 3)$.
Here, the cost $3$ attributed to the additional edge in $G'$ results in the edges $e_1$ and $e_2$ not being cut in the optimum $(0, 0, 0)$ which has cost $0$.
Depicted in (b) is an instance of the minimum cost multicut problem w.r.t.~the graph $G'$ and the same cost function.
Here, the cost $3$ does not prevent the edges $e_1$ and $e_2$ from being cut in the optimum $(1, 1, 0)$ which has cost $-2$.}
\label{figure:lmp}
\end{figure}

As a third and final application of lifted multicuts, we turn to the optimization of graph decompositions by lifted multicuts of minimum cost.

\begin{definition}
\label{definition:lmp}
For any connected graph $G = (V,E)$,
any graph $G' = (V,E')$ with $E \subseteq E'$
and any $c: E' \to \mathbb{R}$,
the instance of the \emph{minimum cost lifted multicut problem}
w.r.t.~$G$, $G'$ and $c$ is the optimization problem
\begin{align}
\min\ \left\{ \sum_{e \in E'} c_e x_e \ \middle|\ x \in X_{GG'} \right\}
    \enspace .
\label{eq:lmp}
\end{align}
\end{definition}
If $E' = E$, \eqref{eq:lmp} specializes to the \emph{minimum cost multicut problem} w.r.t.~$G'$ and $c$ that is also known as \emph{graph partition} or \emph{correlation clustering}.
If $E' \supset E$, 
the minimum cost lifted multicut problem w.r.t.~$G$, $G'$ and $c$ differs from 
the minimum cost multicut problem w.r.t.~$G'$ and $c$.
It has a smaller feasible set $X_{GG'} \subset X_{G'}$, 
as we have shown in Section~\ref{section:lifting}
and depicted for the smallest example in
Fig.~\ref{figure:multicut-polytope} and \ref{figure:lifted-multicut-polytope}.
Unlike the minimum cost multicut problem w.r.t.~$G'$ and $c$, the minimum cost lifted multicut problem w.r.t.~$G$, $G'$ and $c$ is such that any feasible solution $x \in X_{GG'}$ indicates by $x_{vw} = 0$ that the nodes $v$ and $w$ are connected in $G$ by a path of edges labeled $0$.
See also Fig.~\ref{figure:lmp}.
This property can be used to penalize by $c_{vw} > 0$ precisely those decompositions of $G$ for which $v$ and $w$ are in distinct components.
For nodes $v$ and $w$ that are not neighbors in $G$, such costs are sometimes called \emph{non-local attractive}.

To solve instances of the \textsc{apx}-hard minimum cost lifted multicut problem by means of a branch-and-cut algorithm, we study the geometry of lifted multicut polytopes.

\begin{definition}\citep{deza-1997}
\label{definition:multicut-polytope}
For any graph $G = (V,E)$, 
the convex hull $\Xi_G := \conv X_G$ of $X_G$ in $\mathbb{R}^E$
is called the \emph{multicut polytope} of $G$.
\end{definition}

\begin{definition}
\label{definition:lifted-multicut-polytope}
For any connected graph $G = (V ,E)$ and any graph $G' = (V, E')$ with $E \subseteq E'$,
$\Xi_{GG'} := \conv X_{GG'}$ 
is called the \emph{lifted multicut polytope} w.r.t.~$G$ and $G'$.
\end{definition}

Examples are shown in
Fig.~\ref{figure:multicut-polytope} and \ref{figure:lifted-multicut-polytope}, respectively.
In general, the lifted multicut polytope $\Xi_{GG'}$ w.r.t.~graphs $G$ and $G'$ 
(Fig.~\ref{figure:lifted-multicut-polytope})
is a subset of the multicut polytope $\Xi_{G'}$ of the graph $G'$ 
(Fig.~\ref{figure:multicut-polytope}).
By Lemma~\ref{lemma:lifted-multicut}, the system of cycle inequalities \eqref{eq:cycle-inequalities} for $G'$ and cut inequalities \eqref{eq:lmc-cut} for $G$ and $G'$ is redundant as a description of $X_{GG'}$ and thus of $\Xi_{GG'}$.
Below, we study the geometry of $\Xi_{GG'}$.

\subsection{Dimension}

\begin{theorem}
\label{theorem:dimension}
For any connected graph $G = (V,E)$ and any graph $G' = (V,E')$ with $E \subseteq E'$,
$\dim \Xi_{GG'} = |E'|$.
\end{theorem}

\begin{figure}
\ifbool{PRECOMPILED}{
a) \imagetop{\includegraphics[scale=1]{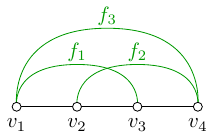}}
\hfill
b) \imagetop{\includegraphics[scale=1]{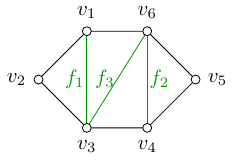}}
}{
a) \imagetop{\begin{tikzpicture}[scale=0.93]
\draw (0, 0) to (3, 0);
\draw[green] plot [smooth, tension=2] coordinates {(0, 0) (1, 0.7) (2, 0)};
\draw[green] plot [smooth, tension=2] coordinates {(1, 0) (2, 0.7) (3, 0)};
\draw[green] plot [smooth, tension=2] coordinates {(0, 0) (1.5, 1.3) (3, 0)};
\node[rectangle, draw=none, fill=none] at (1, 0.9) {$\color{green}f_1$};
\node[rectangle, draw=none, fill=none] at (2, 0.9) {$\color{green}f_2$};
\node[rectangle, draw=none, fill=none] at (1.5, 1.5) {$\color{green}f_3$};
\node[style=vertex, label=below:$v_1$] at (0, 0) {};
\node[style=vertex, label=below:$v_2$] at (1, 0) {};
\node[style=vertex, label=below:$v_3$] at (2, 0) {};
\node[style=vertex, label=below:$v_4$] at (3, 0) {};
\end{tikzpicture}}
b) \imagetop{\begin{tikzpicture}[scale=0.93]
\draw (0.8, 1.6) to (0, 0.8) to (0.8, 0) to (1.8, 0) to (2.6, 0.8) to (1.8, 1.6) to (0.8, 1.6);
\draw[green] (0.8, 0) to (1.8, 1.6);
\draw[green] (0.8, 0) to (0.8, 1.6);
\draw[green] (1.8, 0) to (1.8, 1.6);
\node[style=vertex, label=left:$v_2$] at (0, 0.8) {};
\node[style=vertex, label=below:$v_3$] at (0.8, 0) {};
\node[style=vertex, label=below:$v_4$] at (1.8, 0) {};
\node[style=vertex, label=right:$v_5$] at (2.6, 0.8) {};
\node[style=vertex, label=above:$v_6$] at (1.8, 1.6) {};
\node[style=vertex, label=above:$v_1$] at (0.8, 1.6) {};
\node[rectangle, draw=none, fill=none] at (0.6, 0.8) {$\color{green}f_1$};
\node[rectangle, draw=none, fill=none] at (2, 0.8) {$\color{green}f_2$};
\node[rectangle, draw=none, fill=none] at (1.1, 0.8) {$\color{green}f_3$};
\end{tikzpicture}}
}
\caption{If two nodes $\{v,w\} = f \in F_{GG'}$ are in the same component,
as indicated by $x_f = 0$, this can imply $x_{f'} = 0$ for one or
more $f' \in F \setminus \{f\}$. In (a) $x_{f_3} = 0$ implies
$x_{f_1} = 0$ and $x_{f_2} = 0$. In (b) $x_{f_3} = 0$ implies
$x_{f_1} = 0$ or $x_{f_2} = 0$.}
\label{figure:nullability}
\end{figure}
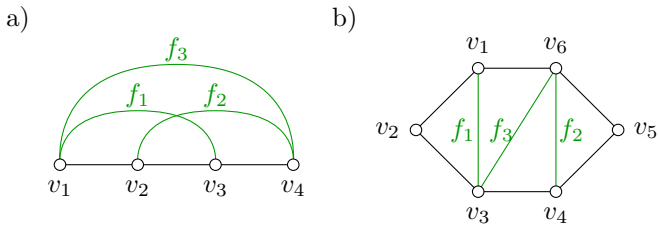

We prove Theorem\,\ref{theorem:dimension} by constructing $|E'| + 1$ multicuts of $G'$ lifted from $G$ whose characteristic functions are affine independent points.
The strategy is to construct, for any $e \in E'$, an $x \in X_{GG'}$ with $x_e = 0$ and ``as many ones as possible''.
The challenge is that edges cannot be labeled independently.
In particular, for $f \in F_{GG'}$, $x_f = 0$ can imply, for certain $f' \in F_{GG'} \setminus \{f\}$, that $x_{f'} = 0$, as illustrated in Fig.~\ref{figure:nullability}.
This structure is made explicit below, in Def.~\ref{definition:hierarchy} and \ref{definition:level} and Lemmata~\ref{lemma:geometry:aux1} and \ref{lemma:geometry:aux2}.

\begin{definition}
\label{definition:hierarchy}
For any connected graph $G = (V,E)$ and any graph $G' = (V,E')$ such that $E \subseteq E'$,
the sequence $(F_n)_{n \in \mathbb{N}}$ of subsets of $F_{GG'}$ defined below 
is called the \emph{hierarchy} of $F_{GG'}$ with respect to $G$:
\begin{enumerate}[(a),nosep]
\item $F_0 = \emptyset$
\item For any $n \in \mathbb{N}$ and any $\{v,w\} = f \in F_{GG'}$: $\{v,w\} \in F_n$ iff there exists a $vw$-path in $G$ such that, for any distinct nodes $v'$ and $w'$ in the path such that $\{v',w'\} \not= \{v,w\}$, either $\{v',w'\} \not\in F_{GG'}$ or there exists a natural number $j < n$ such that $\{v',w'\} \in F_j$.
\end{enumerate}
\end{definition}

\begin{lemma}
\label{lemma:geometry:aux1}
For any connected graph $G = (V,E)$, 
any graph $G' = (V,E')$ with $E \subseteq E'$
and any $f \in F_{GG'}$, 
there exists an $n \in \mathbb{N}$ such that $f \in F_n$.
\end{lemma}

\begin{definition}
\label{definition:level}
For any connected graph $G = (V,E)$
and any graph $G' = (V,E')$ with $E \subseteq E'$,
the map $\ell: F_{GG'} \to \mathbb{N}$ such that
$\forall f \in F_{GG'} \forall n \in \mathbb{N}: \ell(f) = n \Leftrightarrow f \in F_n \wedge f \not\in F_{n-1}$
is called the \emph{level function} of $F_{GG'}$.
\end{definition}

\begin{lemma}
\label{lemma:geometry:aux2}
For any connected graph $G = (V,E)$, 
any graph $G' = (V,E')$ with $E \subseteq E'$
and for any $f \in F_{GG'}$, 
there exists an $x \in X_{GG'}$, called \emph{$f$-feasible}, such that
\begin{enumerate}[(a),nosep]
\item $x_f = 0$
\item $x_{f'} = 1$ for all $f' \in F_{GG'} \setminus \{f\}$ with $\ell(f') \geq \ell(f)$.
\end{enumerate}
\end{lemma}

\subsection{Facets}

We characterize those edges $e \in E'$ 
for which the inequality $x_e \leq 1$ defines a facet of the lifted multicut polytope 
$\Xi_{GG'}$.

\begin{theorem}
\label{theorem:facets-box-upper}
For any connected graph $G = (V,E)$,
any graph $G' = (V,E')$ with $E \subseteq E'$ 
and any $e \in E'$, 
the inequality $x_e \leq 1$ defines a facet of $\Xi_{GG'}$ 
iff there is no $\{v,w\} = f \in F_{GG'}$ such that $e$ connects a pair of $v$-$w$-cut-vertices\footnote{For any graph $G = (V,E)$ and any $v, w \in V$, a \emph{$v$-$w$-cut-vertex} is a node $u \in V$ that lies on every $vw$-path of $G$.}.
\end{theorem}

Next, we give conditions that contribute to identifying those edges $e \in E'$ 
for which the inequality $0 \leq x_e$ defines a facet of the lifted multicut polytope 
$\Xi_{GG'}$.

\begin{theorem}
\label{theorem:facets-box-lower}
For any connected graph $G = (V,E)$,
any graph $G' = (V,E')$ with $E \subseteq E'$ 
and any $e \in E'$, the following assertions hold:

In case $e \in E$, the inequality $0 \leq x_e$ defines a facet of $\Xi_{GG'}$ iff there is no triangle in $G'$ containing $e$.

In case $uv = e \in F_{GG'}$, the inequality $0 \leq x_e$ defines a facet of $\Xi_{GG'}$ only if the following necessary conditions hold:
\begin{enumerate}[(a),nosep]
\item \label{cond:box-1} There is no triangle in $G'$ containing $e$.
\item \label{cond:box-2} The distance of any pair of $u$-$v$-cut-vertices except $\{u,v\}$ is at least $3$ in $G'$.
\item \label{cond:box-3} There is no triangle of nodes $s, s', t$ in $G'$ such that
$\{s,s'\}$ is a $u$-$v$-separating node set and $t$ is a $u$-$v$-cut-vertex.
\end{enumerate}
\end{theorem}

Next, we characterize those inequalities of 
\eqref{eq:lmc-cycle} and \eqref{eq:lmc-path} 
that are facet-defining for $\Xi_{GG'}$.
\citet{chopra-1993} have shown that an inequality of \eqref{eq:cycle-inequalities} defines a facet of the multicut polytope $\Xi_G$ iff the cycle $C$ is chordless. 
We establish a similar characterization of those inequalities of \eqref{eq:lmc-cycle} and \eqref{eq:lmc-path} that define a facet of the lifted multicut polytope $\Xi_{GG'}$.
For clarity, we introduce some notation:
For any cycle $C$ of $G$ and any $e \in C$, let
\begin{align}
S_{GG'}(e,C) 
    & := \left \{ x \in X_{GG'} \middle \vert x_e = \sum_{\mathclap{e' \in C \setminus \{e\}}} x_{e'} \right \}\\
\Sigma_{GG'}(e,C) 
    & := \conv S_{GG'}(e,C)
\enspace .
\end{align}
For any $vw = f \in F_{GG'}$ and any $vw$-path $P$ in $G$, let
\begin{align}
S_{GG'}(f,P) 
    & := \left \{ x \in X_{GG'} \middle \vert x_{vw} = \sum_{e \in P} x_{e} \right \}\\
\Sigma_{GG'}(f,P) 
    & := \conv S_{GG'}(f,P)
\enspace .
\end{align}

\begin{theorem}
\label{theorem:cycle-facets}
For any connected graph $G = (V,E)$
and any graph $G' = (V,E')$ with $E \subseteq E'$,
the following assertions hold:
\begin{enumerate}[(a),nosep]
\item For any cycle $C$ in $G$ and any $e \in C$, the polytope 
$\Sigma_{GG'}(e,C)$ is a facet of $\Xi_{GG'}$ iff $C$ is chordless in $G'$.
\item For any edge $vw = f \in F_{GG'}$ and any $vw$-path $P$ in $G$, the
polytope $\Sigma_{GG'}(f,P)$ is a facet of $\Xi_{GG'}$ iff
$P \cup \{f\}$ is chordless in $G'$.
\end{enumerate}
\end{theorem}

Inequalities defined by cycles in $G'$ that contain more than one edge from the set $F_{GG'}$ do not occur in \eqref{eq:lmc-cycle} or \eqref{eq:lmc-path}.
They are valid for $\Xi_{GG'}$ as they are valid for $\Xi_{G'} \supseteq \Xi_{GG'}$.
They define a (non-trivial) facet of $\Xi_{GG'}$ only if the cycle is chordless (as chordal cycles are not even facet-defining for $\Xi_{G'}$).
At the same time, chordlessness is not a sufficient condition for facet-definingness of non-trivial cycles.
For example, in Fig.~\ref{figure:nullability}a, the cycle inequality $x_{f_2} \leq x_{f_3} + x_{v_1v_2}$ is dominated by the (non-trivial) valid inequality $x_{f_2} \leq x_{f_3}$.

Next, we consider the cut inequalities \eqref{eq:lmc-cut}.
Examples of cuts that are not facet-defining for $\Xi_{GG'}$ are shown in Fig.~\ref{figure:violated-conditions} in the appendix.
To constrain the class of cuts that are facet-defining, we introduce additional notation:

For any connected graph $G = (V,E)$, any distinct nodes $v, w \in V$ and any $C \in vw\cuts(G)$, we denote by
\begin{align}
G(v, C) & = (V(v,C), E(v,C)) \\
G(w, C) & = (V(w,C), E(w,C))
\end{align}
the largest components of the graph $(V, E \setminus C)$ that contain $v$ and $w$, respectively.
By definition of a $vw\textnormal{-cut}$\footnote{%
    For any graph $G = (V,E)$ and any distinct nodes $v, w \in V$, 
    a $vw\textnormal{-cut}$ of $G$ is a minimal (with respect to inclusion) set $C \subseteq E$ 
    such that $v$ and $w$ are not connected in $(V, E \setminus C)$.
}, we have
\begin{align}
V(v,C) \cap V(w,C) & = \emptyset\\
\wedge \quad V(v,C) \cup V(w,C) & = V
\enspace .
\end{align}

We denote by $F_{GG'}(vw,C)$ the set of those edges in $F_{GG'}$, except $vw$, 
that cross the $vw$-cut $C$ of $G$, i.e.
\begin{align}
F_{GG'}(vw,C) & := \{f \in F_{GG'} \setminus \{vw\} \,|\, f \not\subseteq V(v,C) \, \wedge  \nonumber\\
    & \qquad \quad f \not\subseteq V(w,C)\}
\enspace .
\end{align}

We denote by $G'(vw, C) := (V, F_{GG'}(vw,C)\cup C)$ the subgraph of $G'$ that comprises all edges from $F_{GG'}(vw,C)$ and $C$.
Finally, we define
\begin{align}
S_{GG'}(vw, C)
    & := \left\{
        x \in X_{GG'}
        \middle|
        1 - x_{vw} = \sum_{e \in C} (1 - x_e)
    \right\}
    \label{eq:facet-discrete}\\
\Sigma_{GG'}(vw, C)
    & := \conv S_{GG'}(vw, C) 
\enspace .
\end{align}

\begin{definition}
For any connected graph $G = (V,E)$, 
any distinct $v, w \in V$ and 
any $C \in vw\cuts(G)$,
a component $(V^*, E^*)$ of $G$ is called \emph{properly $(vw, C)$-connected} iff
\begin{align}
v \in V^* 
\ \wedge\  
w \in V^*
\ \wedge\ 
|E^* \cap C| = 1
\enspace .
\end{align}
It is called \emph{improperly $(vw, C)$-connected} iff
\begin{align}
V^* \subseteq V(v,C) 
\ \vee\  
V^* \subseteq V(w,C)
\enspace .
\end{align}
It is called \emph{$(vw, C)$-connected} iff it is properly or improperly $(vw, C)$-connected.
\end{definition}

For any $(vw,C)$-connected component $(V^*,E^*)$ of $G$, we denote by
\begin{align}
F_{V^*} := \{ v'w' = f' \in F_{GG'}(vw,C) \mid v' \in V^* \wedge w' \in V^*\}
\end{align} 
the set of those edges $v'w' = f' \in F_{GG'}(vw,C)$ such that $(V^*,E^*)$ is also $(v'w',C)$-connected.

\begin{theorem}
\label{theorem:facets}
For any connected graph $G = (V, E)$, 
any graph $G' = (V, E')$ with $E \subseteq E'$,
any $vw = f \in F_{GG'}$
and any $C \in vw\cuts(G)$,
$\Sigma_{GG'}(vw, C)$ is a facet of $\Xi_{GG'}$ only if the following necessary conditions hold:
\begin{enumerate}[C1]
\item \label{cond:cut-1} For any $e \in C$, there exists a $(vw,C)$-connected component $(V^*, E^*)$ of $G$ such that $e \in E^*$.

\begin{comment} % old condition 2
\item For any $rs \in F_{GG'}(vw,C)$, 
there exist an $e \in C$ and $(vw,C)$-connected components $(V^*, E^*)$ and $(V^{**}, E^{**})$ of $G$ such that
$e \in E^*$ and $e \in E^{**}$ and $(r \in V^* \wedge s \in V^*)$ and $(r \notin V^{**} \vee s \notin V^{**})$.
\end{comment}

\item \label{cond:cut-2} For any $\emptyset \neq F \subseteq F_{GG'}(vw,C)$, there exists an edge $e \in C$ and $(vw,C)$-connected components $(V^*,E^*)$ and $(V^{**},E^{**})$ of $G$ such that $e \in E^*$ and $e \in E^{**}$ and $\lvert F \cap F_{V^*} \rvert \neq \lvert F \cap F_{V^{**}} \rvert$.

\begin{comment} % old condition 3
\item For any distinct $rs, tu \in F_{GG'}(vw,C)$, 
there exist $(vw,C)$-connected components $(V^*, E^*)$ and $(V^{**}, E^{**})$ of $G$ such that
$(r \in V^* \wedge s \in V^*)$ and $(t \in V^{**} \wedge u \in V^{**})$ and $(r \notin V^{**} \vee s \notin V^{**} \vee t \notin V^* \vee u \notin V^*)$.
\end{comment}

\item \label{cond:cut-3} For any $f' \in F_{GG'}(vw,C)$, any $\emptyset \neq F \subseteq F_{GG'}(vw,C) \setminus \{f'\}$ and any $k \in \mathbb{N}$, there exist $(vw,C)$-connected components $(V^*,E^*)$ and $(V^{**},E^{**})$ with $f' \in F_{V^*}$ and $f' \notin F_{V^{**}}$ such that
\begin{align}
\lvert F \cap F_{V^*} \rvert \neq k \text{ or } \lvert F \cap F_{V^{**}} \rvert \neq 0
\enspace .
\end{align}

\item \label{cond:cut-4} For any $v' \in V(v, C)$, any $w' \in V(w, C)$ and any $v'w'$-path $P = (V_P, E_P)$ in $G'(vw,C)$,
there exists a properly $(vw, C)$-connected component $(V^*, E^*)$ of $G$ such that
\begin{align}
& (v' \notin V^* \vee \exists w'' \in V_P \cap V(w, C): w'' \notin V^*)\nonumber\\
\wedge \quad & (w' \notin V^* \vee \exists v'' \in V_P \cap V(v, C): v'' \notin V^*)
\enspace .
\end{align}

\item \label{cond:cut-5} For any cycle $Y = (V_Y, E_Y)$ in $G'(vw,C)$,
there exists a properly $(vw, C)$-connected component $(V^*, E^*)$ of $G$ such that
\begin{align}
& (\exists v' \in V_Y \cap V(v, C): v' \notin V^*)\nonumber\\
\wedge \quad & (\exists w' \in V_Y \cap V(w, C): w' \notin V^*)
\enspace .
\end{align}
\end{enumerate}
\end{theorem}

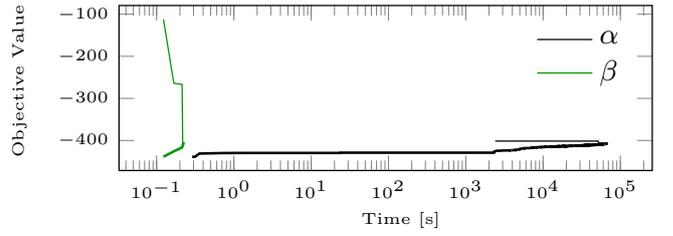
\begin{figure}
\ifbool{PRECOMPILED}{
\includegraphics[width=0.97\columnwidth]{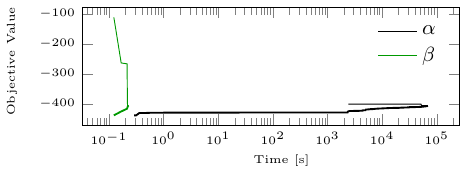}
}{
\begin{tikzpicture}
\small
\begin{axis}[
    width=0.85\columnwidth,
    height=0.4\columnwidth,
    xlabel style={yshift=1.5ex, font=\tiny},
    ylabel style={yshift=-2ex, font=\tiny},
    tick label style={font=\tiny},
    xmode=log,
    xlabel={Time [s]},
    ylabel={Objective Value},
    legend entries={$\alpha$, $\beta$},
    legend style={draw=none, fill=none, nodes={right}, at={(0.9, 0.85)}, anchor=north east}
]
\addplot[color=black] table [col sep=comma, x index=5, y index=1] {ILP-old.txt};
\addplot[color=mycolor1] table [col sep=comma, x index=11, y index=4] {ILP-new.txt};
\addplot[thick, color=black] table [col sep=comma, x index=5, y index=0] {ILP-old.txt};
\addplot[thick, color=mycolor1] table [col sep=comma, x index=11, y index=1] {ILP-new.txt};
\end{axis}
\end{tikzpicture}
}
\caption{Compared above are the separation procedures $\alpha$ (black) and $\beta$ (green) in a branch-and-cut search for solutions of an instance of the minimum cost lifted multicut problem by \citet{keuper-2015a}. 
Upper and lower bounds are depicted as thin and thick lines, respectively.}
\label{figure:algorithm}
\end{figure}

\subsection{Algorithms}

To study the relevance of geometric properties established above, we compare two separation procedures, $\alpha$ and $\beta$, for lifted multicut polytopes.
We implement these for the branch-and-cut algorithm in the software Gurobi.
Our code is available at \url{https://github.com/bjoern-andres/graph}.

The procedure $\alpha$ is canonical and serves as a reference.
It separates infeasible points by any of the inequalities \eqref{eq:lmc-cycle}--\eqref{eq:lmc-cut}.
Violated inequalities of \eqref{eq:lmc-cycle} and \eqref{eq:lmc-path} are found by searching for shortest chordless paths.
Violated inequalities of \eqref{eq:lmc-path} are found by searching for minimum $vw$-cuts.

The procedure $\beta$ is less canonical:
It separates infeasible points by some cycle inequalities w.r.t.~$G'$ (cf.~Theorem~\ref{theorem:cycle-facets}) and by cut inequalities \eqref{eq:lmc-cut}.
Violated cycle inequalities of $G'$ are found by first seaching for paths and cycles as before but then replacing sub-paths by chords in $G'$.
Violated cut-inequalities \eqref{theorem:cycle-facets} are found as before but added to the problem only conditionally:
For each violated inequality of \eqref{eq:lmc-cut} that we find and the corresponding $\{v,w\} \in F_{GG'}$ and $C \in vw\cuts(G)$, we search for a $vw$-path $P$ in $G'$ such that one of the cycle inequalities for the cycle formed by $P$ and $\{v,w\}$ is violated.
If it exists, only the cycle inequality is added.
Otherwise, the cut-inequality is added.
The advantage of $\beta$ over $\alpha$ can be seen in Fig.~\ref{figure:algorithm} for an instance of the minimum cost lifted multicut problem with $|V| = 126$, $|E| = 229$ and $|E'| = 1860$ kindly provided by \citet{keuper-2015a}.

\section{Conclusion}
    By studying the set of all decompositions (clusterings) of a graph through its characterization as a set of lifted multicuts, we have gained three insights:
1.~Toward the definition of classes of decompositions by must-join and must-cut constraints, we have seen that consistency and maximal specificity are \textsc{np}-hard to decide.
This limits unrestricted applications of such constraints in practice.
2.~Toward the comparison of decompositions by metrics, we have defined a generalization of Rand's metric and the boundary metric that enables more detailed analyses of how two decompositions of the same graph differ.
This metric extends to classes of decompositions definable by must-join and must-cut constraints for which it is \textsc{np}-hard to compute.
3.~Toward the optimization of graph decompositions by minimum cost lifted multicuts, we have established some properties of some facets of lifted multicut polytopes. 
These properties have led us to efficient separation procedures and a branch-and-cut algorithm for the minimum cost lifted multicut problem.

\section*{Acknowledgements}
    % !TeX root = 0000.tex

The examples depicted in Fig.~11h and 11j were proposed by Banafsheh Grochulla and Ashkan Mokarian, respectively.
% references
    \bibliographystyle{plainnat}
    \bibliography{0000,andres}
% \clearpage
\appendix
\section*{Appendix}
\section{Multicuts}
    
\paragraph{Proof of Lemma \ref{lemma:characteristic}}
First, we show that for any $\Pi \in D_G$, the image $\phi_G(\Pi)$ is a multicut of $G$. Assume the contrary, i.e.\ there exists a cycle $C$ of $G$ such that $\lvert C \cap \phi_G(\Pi) \rvert = 1$. Let $\{u,v\} = e \in C \cap \phi_G(\Pi)$, then for all $U \in \Pi$ it holds that $u \notin U$ or $v \notin U$. However, $C \setminus \{e\}$ is a sequence of edges $\{w_1,w_2\}, \dotsc, \{w_{k-1},w_k\}$ such that $u = w_1, v = w_k$ and $\{w_i,w_{i+1}\} \notin \phi_G(\Pi)$ for all $1 \leq i \leq k-1$. Consequently, since $\Pi$ is a partition of $V$, there exists some $U \in \Pi$ such that
\begin{align*}
w_1 \in U \wedge w_2 \in U \wedge \dotso \wedge w_{k-1} \in U \wedge w_k \in U.
\end{align*}
This contradicts $w_1 = u \notin U$ or $w_k = v \notin U$.

To show injectivity of $\phi_G$, let $\Pi = \{U_1, \dotsc, U_k\}, \; \Pi' = \{U'_1, \dotsc, U'_\ell\}$ be two decompositions of $G$. Suppose $\Pi \neq \Pi'$, then there exist some $u, v \in V, u \neq v$ and some $U_i \in \Pi$ such that $u, v \in U_i$ and for all $U'_j \in \Pi'$ it holds that $u \notin U'_j$ or $v \notin U'_j$. Thus, $\{u,v\} \in \phi_G(\Pi')$ but $\{u,v\} \notin \phi_G(\Pi)$, which means $\phi_G(\Pi) \neq \phi_G(\Pi')$.

For surjectivity, take some multicut $M \subseteq E$ of $G$. Let $\Pi = \{U_1, \dotsc, U_k\}$ collect the node sets of the connected components of the graph $(V, E \setminus M)$. Apparently, $\Pi$ defines a decomposition of $G$. We have $\{u,v\} \in \phi_G(\Pi)$ if and only if for all $U \in \Pi$ it holds that $v \notin U$ or $u \notin U$. The latter holds true if and only if $\{u,v\}$ is not contained in any connected component of $(V, E \setminus M)$, which is equivalent to $\{v,w\} \in M$. Hence, $\phi_G(\Pi) = M$.

\paragraph{Proof of Lemma \ref{lemma:multicut-complete-graph}}
First, we show that for any $M \in M_{K_V}$ the image $\psi(M)$ is an equivalence relation on $V$. Since $K_V$ is simple, we trivially have $\{v,v\} \notin M$ for any $v \in V$. Therefore, $(v,v) \in \psi(M)$, which means $\psi(M)$ is reflexive. Symmetry of $\psi(M)$ follows from $\{u,v\} = \{v,u\}$ for all $u,v \in V$. Now, suppose $(u,v), (v,w) \in \psi(M)$. Then $\{u,v,\}, \{v,w\} \notin M$ and thus $\{u,w\} \notin M$ (otherwise $C = \{u,v,w\}$ would be a cycle contradicting the definition of a multicut). Hence, $(u,w) \in \psi(M)$, which gives transitivity of $\psi(M)$.

Let $M,M'$ be two multicuts of $K_V$ with $\psi(M) = \psi(M')$. Then
\begin{align*}
\{u,v\} \in M & \iff (u,v) \notin \psi(M) \\
& \iff (u,v) \notin \psi(M') \\
& \iff \{u,v\} \in M'.
\end{align*}
Hence $M = M'$, so $\psi$ is injective.

Let $R$ be an equivalence relation on $V$ and define $M$ by
\begin{align*}
\{u,v\} \in M \iff (u,v) \notin R.
\end{align*}
Transitivity of $R$ implies that $M$ is a multicut of $K_V$. Moreover, by definition, it holds that $\psi(M) = R$. Hence, $\psi$ is also surjective.

\section{Lifted Multicuts}

\paragraph{Proof of Lemma \ref{lemma:lifted-multicut}}
Let $x \in \{0,1\}^{E'}$ be such that $M' = x^{-1}(1)$ is a multicut of $G'$ lifted from $G$. Every cycle in $G$ is a cycle in $G'$. Moreover, for any path $vw = f \in F_{GG'}$ and any $vw$-path $P$ in $G$, it holds that $P \cup \{f\}$ is a cycle in $G'$. Therefore, $x$ satisfies all inequalities \eqref{eq:lmc-cycle} and \eqref{eq:lmc-path}. Assume $x$ violates some inequality of \eqref{eq:lmc-cut}. Then there is an edge $vw \in F_{GG'}$ and some $vw$-cut $C$ in $G$ such that $x_{vw} = 0$ and for all $e \in C$ we have $x_e = 1$. Let $\Pi$ be the partition of $V$ corresponding to $M'$ according to Lemma \ref{lemma:characteristic}. There exists some $U \in \Pi$ with $v \in U$ and $w \in U$. However, for any $uu' = e \in C$ it holds that $u \notin U$ or $u' \notin U$. This means the subgraph $(U,E \cap {U \choose 2})$ is not connected, as $C$ is a $vw$-cut. Hence, $\Pi$ is not a decomposition of $G$, which is a contradiction, because $G$ is connected.

Now, suppose $x \in E'$ satisfies all inequalities \eqref{eq:lmc-cycle}--\eqref{eq:lmc-cut}. We show first that $M ' = x^{-1}(1)$ is a multicut of $G'$. Assume the contrary, then there is a cycle $C'$ in $G'$ and some edge $e'$ such that $C' \cap M' = \{e'\}$. For every $vw = f \in F_{GG'} \cap C' \setminus \{e'\}$ there exists a $vw$-path $P$ in $G$ such that $x_e = 0$ for all $e \in P$. Otherwise there would be some $vw$-cut in $G$ violating \eqref{eq:lmc-cut}, as $G$ is connected. If we replace every such $f$ with its associated path $P$ in $G$, then the resulting cycle violates either \eqref{eq:lmc-cycle} (if $e' \in E$) or \eqref{eq:lmc-path} (if $e' \in F_{GG'}$). Thus, $M'$ is a multicut of $G'$. By connectivity of $G$, the partition $\phi_{G'}^{-1}(M')$ is a decomposition of both $G'$ and $G$. Therefore, $M = \lambda_{GG'}^{-1}(M') = \phi_G(\phi_{G'}^{-1}(M'))$ is a multicut of $G$ and hence $M' = x^{-1}(1)$ is indeed lifted from $G$.

\section{Partial Lifted Multicuts}
    % !TeX root = 0000.tex

\paragraph{Proof of Theorem \ref{theorem:consistency}}

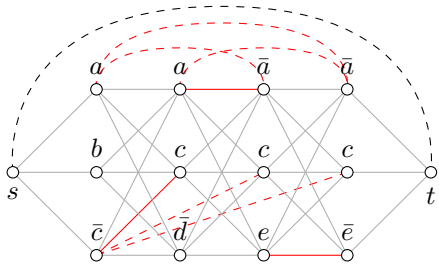
\begin{figure}
\center
\ifbool{PRECOMPILED}{
\includegraphics[scale=1]{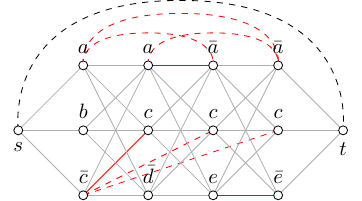}
}{
\begin{tikzpicture}
\draw[undecided] (1, 0) to (3, 0);
\draw [cutcolor] (1, 0) to (2, 1);
\draw [cutcolor,style=dashed] (1, 0) to (4, 1);
\draw[cutcolor,style=dashed] (1, 0) to (3, 1);
\draw [undecided] (2, 1) to (3, 2);
\draw [undecided](3, 0) to (1,2);
\draw[undecided] (1, 2) to (2, 2);
\draw[undecided] (3, 2) to (4, 2);
\draw [undecided](1, 1) to (4, 1);
\draw [undecided](2, 0) to (4, 2);
\draw[undecided] (3, 0) to (4, 1);
\draw[undecided] (1, 1) to (2, 2);
\draw[undecided] (1, 1) to (2, 0);
\draw [undecided](3, 2) to (4, 1);
\draw[undecided] (2, 2) to (4, 0);
\draw [cutcolor] (3,0) to (4,0);
\draw [cutcolor] (2,2) to (3,2);
\draw [undecided](1,0) to (2,2);
\draw [undecided](1,2) to (2,0);
\draw [undecided](2,2) to (3,0);
\draw [undecided](2,0) to (3,2);
\draw [undecided](3,2) to (4,0);
\draw [undecided](3,0) to (4,2);

\draw[undecided] (0, 1) to (1, 0);
\draw[undecided] (0, 1) to (1, 1);
\draw[undecided] (0, 1) to (1, 2);

\draw [undecided](5, 1) to (4, 0);
\draw [undecided](5, 1) to (4, 1);
\draw [undecided](5, 1) to (4, 2);

\draw[joincolor,style=dashed] plot [smooth, tension=2] coordinates {(0, 1) (2.5,3) (5, 1)};

\draw[cutcolor,style=dashed] plot [smooth, tension=2] coordinates {(1, 2) (2,2.5) (3, 2)};
\draw[cutcolor,style=dashed] plot [smooth, tension=2] coordinates {(2, 2) (3,2.5) (4, 2)};
\draw[cutcolor,style=dashed] plot [smooth, tension=2] coordinates {(1, 2) (2.5,2.8) (4, 2)};

\node[style=vertex, label=above:$\bar{c}$] at (1, 0) {};
\node[style=vertex, label=above:$\bar{d}$] at (2, 0) {};
\node[style=vertex, label=above:$e$] at (3, 0) {};
\node[style=vertex, label=above:$\bar{e}$] at (4, 0) {};

\node[style=vertex, label=above:$b$] at (1, 1) {};
\node[style=vertex, label=above:$c$] at (2,1) {};
\node[style=vertex, label=above:$c$] at (3, 1) {};
\node[style=vertex, label=above:$c$] at (4, 1) {};

\node[style=vertex, label=above:$a$] at (1, 2) {};
\node[style=vertex, label=above:$a$] at (2, 2) {};
\node[style=vertex, label=above:$\bar{a}$] at (3, 2) {};
\node[style=vertex, label=above:$\bar{a}$] at (4, 2) {};

\node[style=vertex, label=below:$s$] at (0, 1) {};
\node[style=vertex, label=below:$t$] at (5, 1) {};

\end{tikzpicture} 
}
\caption{To show that the consistency problem is \textsc{np}-hard, we reduce 3-\textsc{sat} to this problem.
Shown above is the instance of the consistency problem constructed for the instance of 3-\textsc{sat} given by the form $(a\vee b\vee \bar{c})\wedge (a\vee c\vee \bar{d})\wedge(\bar{a}\vee c\vee e)\wedge(\bar{a}\vee c\vee \bar{e})$.
Solid and dashed lines depict edges in $E$ and $E' \setminus E$, respectively.
Black means $\tilde x_e = 0$.
Red means $\tilde x_e = 1$.
Grey means $e \notin \dom \tilde x$.}
\label{figure:proof-hardness-consistency}
\end{figure}

Firstly, we show that the consistency problem is in \textsc{np}.
For that, we show that verifying, for any given $x \in \{0,1\}^{E'}$, that $x$ is a completion of $\tilde x$ and a characteristic function of a multicut of $G'$ lifted from $G$ is a problem of polynomial time complexity.
To verify that $x$ is a completion of $\tilde x$, we verify for every $e \in \dom \tilde x$ that $x_e = \tilde x_e$.
This takes time $O(|E|)$.
To verify that $x^{-1}(1)$ is a multicut of $G'$ lifted from $G$, we employ a disjoint set data structure initialized with singleton sets $V$.
For any $\{v,w\} \in x^{-1}(0)$, we call $\union(v,w)$.
Then, we verify for every $\{v,w\} \in x^{-1}(1)$ that $\find(v) \neq \find(w)$.
This takes time $O(|E| + |V| \log |V|)$.
 
To show that the consistency problem is \textsc{np}-hard, we reduce \textsc{3-sat} to this problem.
For that, we consider any instance of \textsc{3-sat} defined by a propositional logic formula $A$ in \textsc{3-sat} form. 
An example is shown in Fig.~\ref{figure:proof-hardness-consistency}.
Let $m$ be the number of variables and $n$ the number of clauses in $A$.

In order to define an instance of the consistency problem w.r.t.~this instance of \textsc{3-sat}, we construct in polynomial time a connected graph $G=(V,E)$, a graph $G'=(V,E')$ with $E \subseteq E'$, and a partial characteristic function $\tilde{x} \in \{0,1,*\}^{E'}$ as described below.
An example of this construction is shown in Fig.~\ref{figure:proof-hardness-consistency}.

\begin{itemize}
\item There are $3n+2$ nodes in $V$. 
Two nodes are denoted by $s$ and $t$.
Additional nodes are organized in $n$ layers.
For $j \in \{1, \ldots, n\}$, the $j$-th layer corresponds to the $j$-th clause in $A$, containing one node for each of the three literals\footnote{A literal is either a variable $a$ or a negated variable $\bar a$.} in the clause.
Every node is labeled with its corresponding literal.
Layer $0$ contains only the node $s$.
Layer $n+1$ contains only the node $t$.

\item Any two consecutive layers connected such that their nodes together induce a complete bipartite subgraph of $G$.
Additionally, any nodes $v$ and $w$ labeled with conflicting literals, $a$ and $\bar a$, that are not already connected in $G$ are connected in $G'$ by an edge $\{v,w\} \in E' \setminus E$.

\item For any edge $\{v,w\} \in E'$ whose nodes $v$ and $w$ are labeled with conflicting literals, we set $\tilde x_{vw} = 1$. 
In addition, we introduce the edge $\{s,t\} \in E' \setminus E$ and define $\tilde x_{st} = 0$.
No other edges are in the domain of $\tilde x$.
\end{itemize}

Observe that $\tilde{x}$ is consistent iff there exists an $st$-path $P$ in $G$ such that no edge or chord $\{v,w\}$ of $P$ is such that $\tilde x_{vw} = 1$.
Any such path is called feasible.
All other paths in $G$ are called infeasible.

Now, we show firstly that the existence of a feasible path implies the existence of a solution to the give instance of \textsc{3-sat}.
Secondly, we show that the existence of a solution to the given instance of \textsc{3-sat} implies the existence of a feasible path.
That suffices.

\begin{enumerate}
\item Let $P$ be a feasible path and let $V_P$ its node set. 
An assignment $\chi$ to the variables of the instance of \textsc{3-sat} is constructed as follows:
For any node $v \in V_P$ whose label is a variable $a$, we define $\chi(a) := \textnormal{true}$. 
For any node $v \in V_P$ whose label is a negated variable $\bar a$, we define $\chi(a) := \textnormal{false}$. 
All remaining variables are assigned arbitrary truth values.
By the properties of $P$, $\chi$ is well-defined and $A[\chi]$ is true.

\item Let $\chi$ be a solution to the given instance of \textsc{3-sat}.
As every clause of $A$ contains one literal that is true, and by construction of $G$, we can choose an $st$-path in $G$ along which all nodes are labeled with literals that are true for the assignment $\chi$.
By virtue of $\chi$ being a solution to the instance of \textsc{3-sat}, any pair of literals that are both true are non-conflicting. 
Thus, $P$ has no edge or chord $\{v,w\}$ such that $\tilde x_{vw} = 1$.
\end{enumerate}

\paragraph{Proof of Lemma \ref{lemma:polynomialConsistency}}

Firstly, suppose that $E\subseteq \dom \tilde x$. 
In this case, it is clear that $\tilde x$ is consistent iff $\tilde x$ satisfies all cycle inequalities \eqref{eq:lmc-cycle} w.r.t.\ the graph $(V,E \cap \dom \tilde x)$. 
This can be checked in time $O(|V| + |E'|)$ as follows:
Label the maximal components of the subgraph $G_{\tilde x}$ of $G$ induced by the edge set $\{e \in E : \tilde x_e = 0\}$. 
Then, for every $\{v,w\} \in E'$ with $\tilde x_{vw} = 1$, check if $v$ and $w$ are in distinct maximal components of $G_{\tilde x}$. 
If so, $\tilde x$ is consistent, otherwise $\tilde x$ is inconsistent.

Now, suppose $\tilde{x}\in \{0,1,*\}^{E'}$ satisfies \eqref{eq:supportedPaths}. 
We show that, similar to the first case, $\tilde{x}$ is consistent iff all inequalities \eqref{eq:lmc-cycle} and \eqref{eq:lmc-path} are satisfied w.r.t.\ the graph $(V,E'\cap \dom \tilde{x})$.
This can be checked analogously to the first case.

Necessity of this condition is clear. 
To show sufficiency, assume this condition holds true. 
We construct some $x \in X_{GG'}[\tilde x]$ as follows. 
For all $e \in \dom \tilde x$, set $x_e := \tilde x_e$. 
For all $\{v,w\} = f \in E' \setminus E$ such that $f \notin \dom \tilde x$ and such that there is a $vw$-path $P$ in $G$ with $\tilde x_e = 0$ for all $e \in P$, set $x_f := 0$. 
For all remaining edges $e$, set $x_e := 1$. 
By construction, $x$ satisfies \eqref{eq:lmc-cycle}, \eqref{eq:lmc-path} and \eqref{eq:lmc-cut}.

\paragraph{Proof of Theorem \ref{theorem:specificity}}

To show that the maximal specificity problem is \textsc{np}-hard, we reduce \textsc{3-sat} to this problem:
For any given instance of \textsc{3-sat} we construct in polynomial time a connected graph $G = (V,E)$, a graph $G' = (V, E')$ with $E \subseteq E'$, and a partial characteristic function $\tilde x \in \{0,1,*\}^{E'}$ as in the proof of Thm.~\ref{theorem:consistency}, except that now, we let $st \notin \dom \tilde x$.

We know that $\tilde x$ is consistent because $\1 \in X_{GG'}[\tilde{x}]$.
We show that $\tilde x$ is maximally specific iff the given instance of \textsc{3-sat} has a solution:

Firstly, every $e \in E' \setminus (\dom \tilde x \cup \{st\})$ is undecided, by the following argument:
(i) There exists an $x \in X_{GG'}[\tilde x]$ with $x_e = 1$, namely $\1$.
(ii) There exists an $x \in X_{GG'}[\tilde x]$ with $x_e = 0$, namely the $x \in \{0,1\}^{E'}$ with $x_e = 0$ and $\forall f \in E' \setminus \{e\}: x_f = 1$. 
To see that $x \in X_{GG'}[\tilde x]$, observe that $e \in E$ and $\tilde x^{-1}(0) = \emptyset$.
Thus, $st$ is the only edge in $E' \setminus \dom \tilde x$ that is possibly decided.
That is:
\begin{align}
E'[\tilde x] \subseteq \{st\} \cup \dom \tilde x
\end{align}

Thus, $\tilde x$ is maximally specific iff $\tilde x$ is undecided.
More specifically, $\tilde x$ is maximally specific iff there exists an $x \in X_{GG'}[\tilde x]$ with $x_{st} = 0$, as we know of the existence of $\1 \in X_{GG'}[\tilde x]$.
Thus, $\tilde x$ is maximally specific iff the given instance of \textsc{3-sat} has a solution, by the arguments made in the proof of Thm.~\ref{theorem:consistency}.

\paragraph{Proof of Lemma~\ref{theorem:specificityPolynomial}}

Observe that $\tilde{x}$ is maximally specific iff $\cl_{GG'} \tilde{x}=\tilde{x}$.
Thus, Lemma~\ref{theorem:specificityPolynomial} follows from Lemma~\ref{lemma:closurePolynomial}.

\paragraph{Proof of Lemma \ref{lemma:partialOrder}}

Reflexivity is obvious. 
Antisymmetry: $(\tilde x\leq \tilde x'\wedge \tilde x'\leq \tilde x)\Rightarrow (\dom \tilde x=\dom \tilde x' \wedge \forall e\in \dom \tilde x: \tilde x_e=\tilde x'_e)$. 
Transitivity: Let $\tilde{x}\leq \tilde{x}'\leq \tilde{x}''$. 
Then $\dom \tilde{x}\subseteq \dom \tilde{x}'\subseteq \dom \tilde{x}''$ and $\forall e\in \dom\tilde{x}: \tilde{x}_e=\tilde{x}'_e=\tilde{x}''_e$.

\paragraph{Proof of Lemma \ref{lemma:maximalSpecificity}}

We show first that $\tilde x'$ is maximal w.r.t.\ $\leq$ in $\tilde X_{GG'}$ iff it is maximally specific. This implies existence and uniqueness of the maximum of $\tilde X_{GG'}[\tilde x]$ by construction via $\dom \tilde x' = E'[\tilde x]$.

Let $\tilde x' \in \tilde X_{GG'}[\tilde x]$ be maximally specific and suppose $\tilde x' \leq \tilde x''$ for some $\tilde x'' \in \tilde X_{GG'}$. Then $\dom \tilde x'' = \dom \tilde x'$, since $X_{GG'}[\tilde x'] \neq X_{GG'}[\tilde x'']$ if $\dom \tilde x'' \setminus E'[\tilde x] \neq \emptyset$. Thus, $\tilde x' = \tilde x''$, which means $\tilde x'$ is maximal w.r.t.\ $\leq$ in $\tilde{X}_{GG'}[\tilde x]$.

Conversely, any maximal element $\tilde x'$ of $\tilde X_{GG'}[\tilde x]$ w.r.t.\ $\leq$ must satisfy $\dom \tilde x' \subseteq E'[\tilde x]$, which means it is maximally specific.

Hence, the unique maximum $\tilde x' \in \tilde X_{GG'}[\tilde x]$ is obtained as follows. For an arbitrary $x \in X_{GG'}[\tilde{x}]$ define $\tilde x'$ via $\tilde x'_e := x_e$ for all decided edges $e \in E'[\tilde x]$.

\paragraph{Proof of Theorem \ref{theorem:closureEquivalence}}

Let us have $\tilde x, \tilde x' \in \tilde X_{GG'}$.
\begin{itemize}
\item The implication $X_{GG'}[\tilde x] = X_{GG'}[\tilde x'] \Rightarrow \tilde X_{GG'}[\tilde x] = \tilde X_{GG'}[\tilde x']:$ follows from the definition of $\tilde X_{GG'} [\tilde{x}]$ in Lemma \ref{lemma:maximalSpecificity}.

\item The implication $\tilde X_{GG'}[\tilde x] = \tilde X_{GG'}[\tilde x'] \Rightarrow \cl_{GG'} \tilde x = \cl_{GG'} \tilde x'$ follows from the definition of the closure of $\tilde{x}$ as the maximum of $\tilde X_{GG'}[\tilde x]$. 

\item The implication $\cl_{GG'} \tilde x = \cl_{GG'} \tilde x'\Rightarrow X_{GG'}[\tilde x] = X_{GG'}[\tilde x']$ follows from 
$\cl_{GG'} \tilde x=\cl_{GG'} \tilde x'\in \tilde{X}_{GG'}[\tilde x]$.
\end{itemize}

\paragraph{Proof of Lemma \ref{lemma:closureLifting}}
Let $x \in X_G$ and define $y = \cl_{GG'} x$. 
Since $\dom x = E$, it holds that $E'[x] = E'$, i.e.\ all edges are decided. 
Therefore, $y^{-1}(1)$ is a multicut of $G'$ and for all $\{v,w\} = f \in E' \setminus E$ it holds that $y_f = 0$ iff there is a $vw$-path $P$ in $G$ such that $x_e = 0$ for all $e \in P$. 
By Lemma \ref{lemma:lifted-multicut}, this implies $y^{-1}(1) = \lambda_{GG'}(x^{-1}(1))$.

\paragraph{Proof of Theorem \ref{theorem:computing-closure}}
Computing closures is at least as hard as deciding maximal specificity: To decide maximal specificity of $\tilde{x}\in \tilde{X}_{GG'}$, compute its closure $\cl_{GG'} \tilde x$. Then $\tilde x$ is maximally specific iff $\dom \tilde x = \dom \cl_{GG'} \tilde x$, i.e., if $\tilde x = \cl_{GG'} \tilde x$. 
By Theorem \ref{theorem:specificity}, this means computing closures is \textsc{np}-hard.

\paragraph{Proof of Lemma \ref{lemma:closurePolynomial}}
Let $\tilde x \in \tilde X_{GG'}$ and $\tilde y = \cl_{GG'} \tilde x$.

Suppose first that $E=E'$. 
We describe how to compute $\tilde y$ efficiently. 
Obviously, we must set $\tilde y_e = \tilde x_e$ for all $e \in \dom \tilde x$. 
Furthermore, we must set $\tilde y_{vw} = 0$ for all $\{v,w\} \in E \setminus \dom \tilde x$ such that there is a $vw$-path $P$ in $G$ with $\tilde x_e=0$ for all $e \in P$.
Moreover, we must set $\tilde y_{vw} = 1$ for all $\{v,w\} \in E \setminus \dom\tilde x$ that satisfy
\begin{align}
& \exists P \in vw\text{-paths}(G) \ 
\exists! e\in P: \nonumber\\
& \qquad \tilde x_e=1 \ \wedge\ \forall e'\in P \setminus \{e\}: \tilde x_{e'}=0
\enspace .
\end{align}
Therefor, initialize a disjoint-set data structure with singleton sets $V$.
Apply the union operation on all edges $e\in \dom \tilde x$ where $\tilde x_e=0$, i.e.\ \emph{contract} all $0$-labeled edges. 
Then, set $\tilde y_e = 0$ for all edges that connect nodes of the same component. 
If there is an edge $e'$ between two components such that $\tilde x_{e'} = 1$, then for all edges $e$ between those components set $\tilde y_e=1$. 
The remaining edges are undecided by $\tilde x$. 
In case we only want to decide maximal specificity, we can stop upon finding the first edge $e\in \dom \tilde y \setminus \dom \tilde x$.

Now suppose that $E\subseteq \dom \tilde x$. 
In this case, all edges are decided, because $\tilde{x}|_E\in X_G$. 
According to Lemma~\ref{lemma:closureLifting}, the closure $\tilde y$ corresponds to the lifting of $\tilde{x}|_E$ to $G'$. 
Therefore, to obtain $\tilde y$, compute the decomposition of $G$ associated to $\tilde{x}|_E$ using, e.g., a disjoint-set data structure.
Set $\tilde y_e = 0$ if $e$ is an edge within a component.
Set $\tilde y_e=1$ if $e$ is an edge between components.

\section{Metrics}
    % !TeX root = 0000.tex

\paragraph{Proof of Theorem \ref{theorem:metric}}

Symmetry and non-negativity follow directly from the definition, and so does $d^\mu_{E'}(x,x)=0$ for all $x \in X_{GG''}$.
For any $e \in E'$, the form $d^1_{e}$ on $E' \times E'$ is a Hamming metric on words from the alphabet $\{0,1\}$.
Therefore, it satisfies the triangle inequality. 
Hence, for any $x, y , z \in X_{GG''}$:
\begin{align}
d^\mu_{E'}(x,z) & = \sum_{e \in E'} \mu_e d^1_e(x,z) \\
& \leq \sum_{e \in E'} \mu_e (d^1_e(x,y) + d^1_e(y,z)) \\
& = \sum_{e \in E'} \mu_e d^1_e(x,y) + \sum_{e \in E'} \mu_e d^1_e(y,z) \\
& = d^\mu_{E'}(x,y) + d^\mu_{E'}(y,z),
\end{align}
Thus, $d^\mu_{E'}$ is a pseudo-metric on $X_{GG''}$.

If $E\subseteq E'$, then $G'=G''$ and thus, $X_{GG''}=X_{GG'} \subseteq X_{G'}$. 
For any two $x,x'\in X_{GG''}\subseteq X_{G'}$, it holds that $d^\mu_{E'}(x,x')=0$ iff $d^1_e(x,x') = 0$ for all $e \in E'$, i.e.\ iff $x=x'$.
Conversely, suppose there exists some $e \in E \setminus E'$. Define $x, x' \in X_{GG''}$ via $x_{e'} = x'_{e'} = 1$ for all $e' \in E'' \setminus \{e\}$ and $x_e = 1$, $x'_e = 0$. It holds that $x\neq x'$ but $d^\mu_{E'}(x,x')=0$.

\paragraph{Proof of Theorem \ref{theorem:metric-partial}}

\begin{table}[]
\centering
\caption{The left- and right-hand side of the inequality $\theta d^1_e(\tilde x, \tilde z) \leq \theta d^1_e(\tilde x, \tilde y) + \theta d^1_e(\tilde y, \tilde z)$ for all possible combinations of values $\tilde x_e, \tilde y_e, \tilde z_e$ where $\tilde x, \tilde y, \tilde z \in \hat X_{GG'}$.  The right-hand side is always greater or equal the left-hand side iff $0.5 \leq \theta$.}\label{tab:triangle-values}
\def\arraystretch{0.8}
\begin{tabular}{ccccc}
\toprule
$\tilde x_e$ & $\tilde y_e$ & $\tilde z_e$ & lhs & rhs \\
\midrule
0 & 0 & 0 & 0 & 0 \\
1 & 1 & 1 & 0 & 0 \\
* & * & * & 0 & 0 \\
0 & * & 1 & 1 & $2\theta$ \\
0 & 1/0 & 1 & 1 & 1 \\
0 & 0/* & * & $\theta$ & $\theta$ \\
1 & 1/* & * & $\theta$ & $\theta$ \\
\bottomrule
\end{tabular}
\end{table}

We first prove that $\tilde d^\theta_{E'}$ is a metric on $\hat X_{GG'}$. 
For any $\tilde{x}\in \hat{X}_{GG'}$, we have $\cl_{GG'} \tilde{x}=\tilde{x}$.
Thus, for all $\tilde{x},\tilde{x}' \in \hat{X}_{GG'}$, we have $\tilde d^\theta_{E'}(\tilde x, \tilde x') = d^\theta_{E'}(\tilde x, \tilde x')$.  
Therefore, positive definiteness and symmetry are obvious from the definition of $d^\theta_{E'}(\tilde x, \tilde x')$. 
To establish the triangle inequality for $d^\theta_{E'}$, we prove it for $\theta d^1_e$ and any edge $e \in E'$.
Let $\tilde x, \tilde y, \tilde z \in \hat X_{GG'}$ and consider the inequality
\begin{align}
\theta d^1_e(\tilde x, \tilde z) \leq \theta d^1_e(\tilde x, \tilde y) + \theta d^1_e(\tilde y, \tilde z). \label{eq:metric-triangle-ineq}
\end{align}
In Tab.~\ref{tab:triangle-values}, the left-hand side and right-hand side of \eqref{eq:metric-triangle-ineq} are evaluated for all possible assignments of values to $\tilde x_e, \tilde y_e, \tilde z_e$. 
It is apparent form this table that \eqref{eq:metric-triangle-ineq} holds iff $\theta\geq 0.5$.

We now show that $\tilde d^\theta_{E'}$ is a pseudo-metric on $\tilde{X}_{GG'}$. 
Symmetry and non-negativity are obvious from the definition. 
For all $\tilde{x}\in \tilde{X}_{GG'}$, we have $\tilde d^\theta_{E'}(\tilde{x},\tilde{x})=0$. 
Since $\tilde d^\theta_{E'}(\tilde x, \tilde x') = \tilde d^\theta_{E'}(\cl_{GG'} \tilde x, \cl_{GG'} \tilde x')$ and $\cl_{GG'} \tilde x \in \hat{X}_{GG'}$ for any $\tilde x \in \tilde X_{GG'}$, the triangle inequality follows from the fact that $\tilde d^\theta_{E'}$ is a metric on $\hat{X}_{GG'}$.

Finally, it holds that $\tilde d^\theta_{E'}(\tilde x, \tilde x') = 0$ iff $\cl_{GG'} \tilde x= \cl_{GG'} \tilde x'$, which in turn is equivalent to $\tilde X_{GG'}[\tilde x] = \tilde X_{GG'}[\tilde x']$, by Theorem \ref{theorem:closureEquivalence}. 
This proves property \eqref{eq:invariance}.

\section{Polyhedral Optimization}
    % !TeX root = 0000.tex

\paragraph{Proof of Theorem\,\ref{theorem:dimension}}
The all-one vector $\1 \in \{0,1\}^{E'}$ is such that $\1 \in X_{GG'}$.

For any $e \in E$, $x^e \in \{0,1\}^{E'}$ such that
$x^e_e = 0$ and $x^e_{E \setminus \{e\}} = 1$ and $x^e_{F_{GG'}} = 1$
holds $x^e \in X_{GG'}$.

For any $f \in F_{GG'}$, any $f$-feasible $x^f \in \{0,1\}^{E'}$ is such that $x^f \in X_{GG'}$.
Moreover, $x^f$ can be chosen such that one shortest path connecting the two nodes
in $f$ is the only component containing more than one node.

For any $e \in E$, let $y^e \in \mathbb{R}^{E'}$ such that
\begin{equation}
y^e = \1 - x^e \enspace .
\end{equation}

For any $f \in F_1$,
choose an $f$-feasible $x^f$ and let $y^f \in \mathbb{R}^{E'}$ such that
\begin{equation}
y^f = \1 - x^f - \hspace{-3ex} \sum_{\{e \in E | x^f_e = 0\}} \hspace{-3ex} y^e \enspace .
\end{equation}

For any $n \in \mathbb{N}$ such that $n > 1$ and any $f \in F_n$,
choose an $f$-feasible $x^f$ and let $y^f \in \mathbb{R}^{E'}$ such that
\begin{equation}
y^f = \1 - x^f \hspace{2ex}
- \hspace{-7ex} \sum_{\{f' \in F_{GG'} | f' \not= f \wedge x^f_{f'} = 0\}} \hspace{-6ex} y^{f'} \hspace{6ex}
- \hspace{-3ex} \sum_{\{e \in E | x^f_e = 0\}} \hspace{-3ex} y^e \enspace .
\end{equation}

Here, $\ell(f') < \ell(f) \leq n$, by definition of $f$-feasibility.
Thus, all $y^{f'}$ are well-defined by induction (over $n$).

Observe that $\{y^e\ |\ e \in E'\}$ is the unit basis in $\mathbb{R}^{E'}$. Moreover, each of its elements is a linear combination of 
$\{ \1 - x^e\ |\ e \in E'\}$
which is therefore linearly independent.

Thus, $\{\1\} \cup \{x^e\ |\ e \in E'\}$ is affine independent.
It is also a subset of $X_{GG'}$ and, therefore, a subset of $\Xi_{GG'}$.
Thus, $\dim \Xi_{GG'} = |E'|$.

\paragraph{Proof of Lemma \ref{lemma:geometry:aux1}}
Let $\{v,w\} = f \in F_{GG'}$ and 
let $d(v,w)$ the length of a shortest $vw$-path in $G$.
Then, $d(v,w) > 1$ because $F_{GG'} \cap E = \emptyset$.

If $d(v,w) = 2$, there exists a $u \in V$ such that 
$\{v,u\} \in E$ and $\{u,w\} \in E$. 
Moreover, $\{v,u\} \notin F_{GG'}$ and $\{u,w\} \notin F_{GG'}$, as $F_{GG'} \cap E = \emptyset$.
Thus, $f \in F_1$.

If $d(v,w) = m$ with $m > 2$, 
consider any shortest $vw$-path $P$ in $G$.
Moreover, let $F' \subseteq F_{GG'}$ such that, 
for any $\{v', w'\} = f' \in F_{GG'}$,
$f' \in F'$ iff $v' \in P$ and $w' \in P$ and $f' \neq f$.
If $F' = \emptyset$ then $f \in F_1$.
Otherwise:
\begin{align}
\forall \{v', w'\} \in F': \quad d(v', w') < m
\end{align}
and thus:
\begin{align}
\forall f' \in F'\ \exists n_{f'} \in \mathbb{N}: \quad f' \in F_{n_{f'}}
\end{align}
by induction (over $m$).
Let
\begin{align}
n = \max_{f' \in F'} n_{f'} \enspace .
\end{align}
Then, $f \in F_{n+1}$.

\paragraph{Proof of Lemma \ref{lemma:geometry:aux2}}
For any $\{v,w\} = f \in F_{GG'}$, let $P$ be a shortest $vw$-path in $G$
and let 
\begin{align}
F_{GG'}' & := \{\{v', w'\} \in F_{GG'} \ |\ v' \in P \wedge w' \in P\}\\
F_{GG'}'' & := F_{GG'} \setminus F_{GG'}' 
\enspace .
\end{align}
Moreover, let $x \in \{0,1\}^{E'}$ with 
$x_P = 0$ and
$x_{E \setminus P} = 1$ and 
$x_{F_{GG'}'} = 0$ and 
$x_{F_{GG'}''} = 1$.
$P$ has no chord in $E$,
    because it is a shortest path.
Thus, $x \in X_{GG'}$.

\paragraph{Proof of Theorem\, \ref{theorem:facets-box-upper}}
Let $S = \{x \in X_{GG'} \mid x_e=1\}$ and put $\Sigma = \conv S$. 

To show necessity, suppose there is some $vw = f \in F_{GG'}$ such that $e$ connects a pair of $v$-$w$-cut-vertices. Then, for any $vw$-path $P$ in $G$, either $e \in P$ or $e$ is a chord of $P$. We claim that we have $x_f=1$ for any $x \in S$. This gives $\dim \Sigma \leq \lvert E' \rvert - 2$, so the inequality $x_e \leq 1$ cannot define a facet of $\Xi_{GG'}$. If there are no $vw$-paths that have $e$ as a chord, then $\{e\}$ is a $vw$-cut and the claim follows from the corresponding inequality of \eqref{eq:lmc-cut}. Otherwise, every $vw$-path $P$ that has $e$ as a chord contains a subpath $P'$ such that $P' \cup \{e\}$ is a cycle. Thus, for any $x \in S$, the inequalities \eqref{eq:lmc-cycle} or \eqref{eq:lmc-path} (for $e \in E$ or $e \in F_{GG'}$, respectively) imply the existence of some $e_{P'} \in P'$ such that $x_{e_{P'}} = 1$.
Let $\mathcal{P}$ denote the set of all such paths $P'$. Apparently, the collection $\bigcup_{P' \in \mathcal{P}} \{e_{P'}\} \cup \{e\}$ is a $v$-$w$-separating set of edges. Therefore, it contains some subset $C$ that is a $vw$-cut. This gives $x_f = 1$ via the inequality of \eqref{eq:lmc-cut} corresponding to $C$.

We turn to the proof of sufficiency. Assume there is no $vw = f \in F_{GG'}$ such that $e$ connects a pair of $v$-$w$-cut-vertices in $G$. The construction of an affine independent $\lvert E' \rvert$-element-subset of $S \subset X_{GG'}$ is analogous to the proof of Theorem \ref{theorem:dimension}. 
The assumption guarantees for any $f \in F_{GG'}$ with $f \neq e$ the existence of an $f$-feasible $x \in S$ such that there is a $vw$-path $P$ with $x_P = 0$. In particular, the hierarchy on $F_{GG'}$ defined by the level function $\ell$ remains unchanged (if $e \in F_{GG'}$, then $\ell(e) \geq \ell(f)$ for all $f \in F_{GG'}$).
Hence, $\dim \Sigma = \lvert E' \rvert - 1$, which means $\Sigma$ is a facet of $\Xi_{GG'}$.

\paragraph{Proof of Theorem\, \ref{theorem:facets-box-lower}}
Let $S = \{x \in X_{GG'} \mid x_e=0\}$ and put $\Sigma = \conv S$.

Consider the case that $e \in E$. 
Let $G_{[e]}$ and $G'_{[e]}$ be the graphs obtained from $G$ and $G'$, respectively, by contracting the edge $e$. 
The lifted multicuts $x^{-1}(1)$ for $x \in S$ correspond bijectively to the multicuts of $G'_{[e]}$ lifted from $G_{[e]}$. 
This implies $\dim \Sigma = \dim \Xi_{G_{[e]} G'_{[e]}}$. 
The claim follows from Theorem \ref{theorem:dimension} and the fact that $G'_{[e]}$ has $\lvert E' \rvert - 1$ many edges if and only if $e$ is not contained in any triangle in $G'$.

Now, suppose $uv = e \in F_{GG'}$. We show necessity of Conditions \ref{cond:box-1}-\ref{cond:box-3} by proving that if any of them is violated, then all $x \in S$ satisfy some additional, orthogonal equality and thus, $\dim \Sigma \leq \lvert E' \rvert - 2$.

First, assume that \ref{cond:box-1} is violated. 
Hence, there are edges $e',e'' \in E'$ such that $T = \{e,e',e''\}$ is a triangle in $G'$. 
Every $x \in S$ satisfies the cycle inequalities
\begin{align}
x_{e'} & \leq x_e + x_{e''} \label{eq:proof-facets-box-lower-1} \\
x_{e''} & \leq x_e + x_{e'} \label{eq:proof-facets-box-lower-2}
\end{align}
by Lemma~\ref{lemma:encoding} applied to the multicut $x^{-1}(1)$ of $G'$.
Every $x \in S$ satisfies $x_{e'} = x_{e''}$, by \eqref{eq:proof-facets-box-lower-1} and \eqref{eq:proof-facets-box-lower-2} and $x_e = 0$.

Next, assume that \ref{cond:box-2} is violated. 
Consider a violating pair $\{u',v'\} \neq \{u,v\}, u' \neq v'$ of $u$-$v$-cut-vertices. 
For every $x \in S$, there exists a $uv$-path $P$ in $G$ with $x_P = 0$, as $x_e = 0$.
Any such path $P$ has a sub-path $P'$ from $u'$ to $v'$ because $u'$ and $v'$ are $u$-$v$-cut-vertices.
\begin{itemize}
\item If the distance of $u'$ and $v'$ in $G'$ is 1, then $u'v' \in E'$.
If $u'v' \in P$, then $x_{u'v'} = 0$ because $x_P = 0$.
Otherwise, $x_{u'v'} = 0$ by $x_{P'} = 0$ and the cycle/path inequality
\begin{align}
x_{u'v'} \leq \sum_{\hat e \in P'} x_{\hat e}
\enspace .
\end{align}
Thus $x_{u'v'} = 0$ for all $x \in S$.
\item If the distance of $u'$ and $v'$ in $G'$ is 2, there is a $u'v'$-path $P''$ in $G'$ consisting of two distinct edges $e',e'' \in E'$.
We show that all $x \in S$ satisfy $x_{e'} = x_{e''}$:
\begin{itemize}
\item If $e' \in P$ and $e'' \in P$ then $x_{e'} = x_{e''} = 0$ because $x_P = 0$.
\item If $e' \in P$ and $e'' \notin P$ then $x_{e'} = x_{e''} = 0$ by $x_P = 0$ and the cycle/path inequality
\begin{align}
x_{e''} \leq \sum_{\hat e \in P' \setminus \{e'\}} x_{\hat e}
\enspace .
\end{align}
\item If $e' \notin P$ and $e'' \notin P$  then $x_{e'} = x_{e''}$ by $x_P = 0$ and the cycle/path inequalities
\begin{align}
x_{e''} & \leq x_{e'} + \sum_{\hat e \in P'} x_{\hat e} \\
x_{e'} & \leq x_{e''} + \sum_{\hat e \in P'} x_{\hat e} 
\enspace .
\end{align}
\end{itemize}
\end{itemize}

Now, assume that \ref{cond:box-3} is violated.
Hence, there exists a $u$-$v$-cut-vertex $t$ and a $u$-$v$-separating set of vertices $\{s,s'\}$ such that $\{ts,ts',ss'\}$ is a triangle in $G'$. We have that all $x \in S$ satisfy $x_{ss'} = x_{ts} + x_{ts'}$ as follows. At most one of 
$x_{ts}$ and $x_{ts'}$ is $1$, because $t$ is a $u$-$v$-cut-vertex and $\{s,s'\}$ is $u$-$v$-separating as well. Moreover, $x_{ts} + x_{ts'} = 0$ if and only if $x_{ss'} = 0$.

\paragraph{Proof of Theorem\, \ref{theorem:cycle-facets}}
Note that both $C$ and $P \cup \{f\}$ are cycles in $G'$. 
We show that, for any chordal cycle $C'$ of $G'$ and any $e \in C'$, the inequality
\begin{align}
x_e \leq \sum_{e' \in C' \setminus \{e\}} x_{e'} \label{eq:mc-cycle}
\end{align}
is not facet-defining for $\Xi_{G'}$. 
This implies that \eqref{eq:mc-cycle} cannot be facet-defining for $\Xi_{GG'}$ either,
as $\Xi_{GG'} \subseteq \Xi_{G'}$ and $\dim \Xi_{GG'} = \dim \Xi_{G'}$. 
Hence, for facet-definingness of \eqref{eq:lmc-cycle} and \eqref{eq:lmc-path}, it is necessary that $C$ and $P \cup \{f\}$ be chordless in $G'$.

For this purpose, consider some cycle $C'$ of $G'$ with a chord $uv = e' \in E'$. We may write $C' = P_1 \cup P_2$ where $P_1$ and $P_2$ are edge-disjoint $uv$-paths such that $C_1 = P_1 \cup \{e'\}$ and $C_2 = P_2 \cup \{e'\}$ are cycles in $G'$. Let $e \in C'$, then either $e \in P_1$ or $e \in P_2$. W.l.o.g.\ we may assume $e \in P_1$. The inequalities
\begin{align}
x_e \leq \sum_{e'' \in C_1 \setminus \{e\}} x_{e''}, \label{eq:mc-cycle-1} \\
x_{e'} \leq \sum_{e'' \in C_2 \setminus \{e'\}} x_{e''} \label{eq:mc-cycle-2}
\end{align}
are both valid for $\Xi_{G'}$. Moreover, since $e' \in C_1$, \eqref{eq:mc-cycle-1} and \eqref{eq:mc-cycle-2} imply \eqref{eq:mc-cycle} via
\begin{align}
x_e & \leq \sum_{e'' \in C_1 \setminus \{e\}} x_{e''} = \sum_{e'' \in C_1 \setminus \{e,e'\}} x_{e''} + x_{e'} \nonumber\\
& \leq \sum_{e'' \in C_1 \setminus \{e,e'\}} x_{e''} + \sum_{e'' \in C_2 \setminus \{e'\}} x_{e''} \nonumber\\
& = \sum_{e'' \in C' \setminus \{e\}} x_{e''}.
\end{align}
Thus, \eqref{eq:mc-cycle} is not facet-defining for $\Xi_{G'}$.

For the proof of sufficiency, suppose the cycle $C$ of $G$ is chordless in $G'$ and let $e \in C$. Let $\Sigma$ be a facet of $\Xi_{GG'}$ such that $\Sigma_{GG'}(e,C) \subseteq \Sigma$ and suppose it is induced by the inequality
\begin{align}
\sum_{e' \in E'} a_{e'} x_{e'} \leq \alpha \label{eq:facet-def-ineq}
\end{align}
with $a \in \mathbb{R}^{E'}$ and $\alpha \in \mathbb{R}$,
i.e., $\Sigma = \conv S$, where
\begin{align}
S := \left\{ x \in X_{GG'} \,\middle|\, \sum_{e' \in E'} a_{e'} x_{e'} = \alpha \right\}.
\end{align}
For convenience, we also define the linear space 
\begin{align}
L := \left\{ x \in \mathbb{R}^{E'} \,\middle|\, \sum_{e' \in E'} a_{e'} x_{e'} = \alpha \right\}.
\end{align}

As $0 \in S_{GG'}(e,C) \subseteq S$, we have $\alpha = 0$. We show that \eqref{eq:facet-def-ineq} is a scalar multiple of \eqref{eq:lmc-cycle} and thus $\Sigma_{GG'}(e,C) = \Sigma$.

Let $y \in \{0,1\}^{E'}$ be defined by
\begin{align}
y_C = 0, \quad y_{E'\setminus C} = 1,
\end{align}
i.e.\ all edges except $C$ are cut. Then $y \in S_{GG'}(e,C) \subseteq S$, since $C$ is chordless. 

For any $e' \in C \setminus \{e\}$, the vector $x \in \{0,1\}^{E'}$ with
\begin{align}
x_{C \setminus \{e,e'\}} = 0, \quad x_{E' \setminus C \cup \{e,e'\}} = 1
\end{align}
holds $x \in S_{GG'}(e,C) \subseteq S$. Therefore, $y-x \in L$. Thus,
\begin{align}
\forall e' \in C \setminus \{e\}: \quad
	a_{e'} = -a_e 
\enspace. 
\label{eq:proof-cycle-facets-1}
\end{align}

It remains to show that $a_{e'} = 0$ for all edges $e' \in E' \setminus C$. 
We proceed by considering edges from $E$ and $F_{GG'}$ separately. 
% For the remainder of the proof we write $e' = uv$ for nodes $u,v \in V$. 
We consider the nodes $u,v \in V$ such that $uv = e'$.
W.l.o.g., we assume that $v$ does not belong to $C$.
This is possible because $C$ does not have a chord in $G'$.

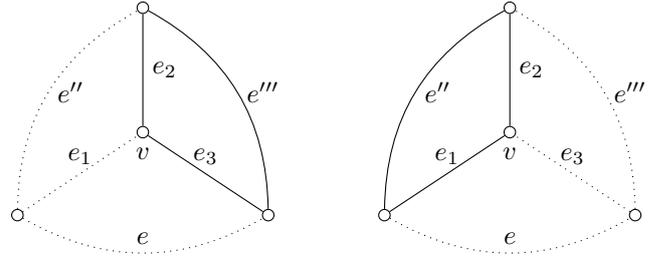
\begin{figure}
\ifbool{PRECOMPILED}{
{\ } \imagetop{\includegraphics[scale=1]{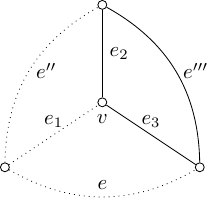}}
\hfill
{\ } \imagetop{\includegraphics[scale=1]{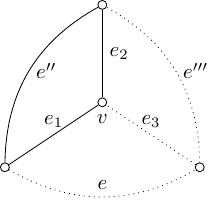}}
}{
{\ } \imagetop{\begin{tikzpicture}[xscale=1]
\node (1) [style=vertex,fill=white] at (0, 0) {};
\node (2) [style=vertex,fill=white] at (3, 0) {};
\node (3) [style=vertex,fill=white] at (1.5, 2.5) {};
\node (4) [style=vertex,fill=white,label=below:$v$] at (1.5, 1) {};

\draw[style=cut-edge] (1) edge[bend right] node[above] {$e$} (2);
\draw[-] (2) edge[bend right] node[right] {$e'''$} (3);
\draw[style=cut-edge] (1) edge[bend left] node[right] {$e''$} (3);
\draw (3) -- node[right] {$e_2$} (4);
\draw[style=cut-edge] (1) edge node[above] {$e_1$} (4);
\draw[-] (2) edge node[above] {$e_3$} (4);
\end{tikzpicture}}
\hfill
{\ } \imagetop{\begin{tikzpicture}[xscale=1]
\node (1) [style=vertex,fill=white] at (0, 0) {};
\node (2) [style=vertex,fill=white] at (3, 0) {};
\node (3) [style=vertex,fill=white] at (1.5, 2.5) {};
\node (4) [style=vertex,fill=white,label=below:$v$] at (1.5, 1) {};

\draw[style=cut-edge] (1) edge[bend right] node[above] {$e$} (2);
\draw[-,style=cut-edge] (2) edge[bend right] node[right] {$e'''$} (3);
\draw[-] (1) edge[bend left] node[right] {$e''$} (3);
\draw[-] (3) edge node[right] {$e_2$} (4);
\draw[-] (1) edge node[above] {$e_1$} (4);
\draw[style=cut-edge] (2) edge node[above] {$e_3$} (4);
\end{tikzpicture}}
}
\caption{The figure illustrates the argument from case \ref{case:proof-cycle-facets-2} in the proof of Theorem \ref{theorem:cycle-facets} for the cycle $C = \{e,e'',e'''\}$. In this example, $e_3 = e'$, $e_1 \in F_{GG'}$ and $e_2 \in E$. The left multicut is chosen for $i=1$ and the right one for $i=2$.}
\label{figure:proof-cycle-facets-1}
\end{figure}

Firstly, consider $e' \in E$ and distinguish the following cases:
\begin{enumerate}[(i)]
\item \label{case:proof-cycle-facets-1} If $e'$ connects two nodes not contained in $C$ or it is the only edge connecting some node in $C$ to $v$, then for $x \in \{0,1\}^{E'}$, defined by
\begin{align}
x_C = 0, \quad x_{e'} = 0, \quad x_{E' \setminus (C \cup \{e'\})} = 1,
\end{align}
it holds that $x \in S_{GG'}(e,C) \subseteq S$. Therefore, $y-x \in L$, which evaluates to $a_{e'} = 0$.
\item \label{case:proof-cycle-facets-2} Otherwise, let $E'_{C,v} := \{ \{u',v\} \in E' \mid u' \text{ belongs to } C\}$ denote the set of edges in $E'$ that connect $v$ to some node in $C$. By assumption, we have that $\lvert E'_{C,v} \rvert \geq 2$. Now, pick some direction on $C$ and traverse $C$ from one endpoint of $e$ to the other endpoint of $e$. We may order the edges $E'_{C,v} = \{e_1, \dotsc, e_k\}$ such that the endpoint of $e_i$ appears before the endpoint of $e_{i+1}$ in the traversal of $C$. We show that $a_{e_i} = 0$ for all $1 \leq i \leq k$: 

For the vector $x \in \{0,1\}^{E'}$ defined by
\begin{align}
x_{e''} = \begin{cases} 0 & \text{if } e'' \in C \\ 0 & \text{if } e'' \in E'_{C,v} \\ 1 & \text{else}, \end{cases}
\end{align}
it holds that $x \in S_{GG'}(e,C) \subseteq S$. Therefore, $y-x \in L$. Thus:
\begin{align}
\sum_{1 \leq i \leq k} a_{e_i} = 0. \label{eq:proof-cycle-facets-2}
\end{align}

Consider the $m \in \{1, \dotsc, k \}$ such that $e' = e_m$. 
For any $i$ with $1 \leq i \leq m-1$, consider the following construction that is illustrated also in Fig.~\ref{figure:proof-cycle-facets-1}:
Let $e'' \in C$ be some edge between the endpoints of $e_i$ and $e_{i+1}$. 
%
% Cut both $e$ and $e''$. 
% Additionally, if $e_i \in E$, then cut all edges $e_j$ for $j > i$. 
% Otherwise, cut all $e_j$ for $j \leq i$. 
% See Figure \ref{figure:proof-cycle-facets-1} for an illustration. 
% In other words, define $x \in \{0,1\}^{E'}$ via
%
If $e_i \in E$, define $x \in \{0,1\}^{E'}$ via
\begin{align}
& x_e = x_{e''} = 1 \\
& x_{C \setminus \{e,e''\}} = 0 \\
\forall j \leq i: \quad & x_{e_j} = 0 \\
\forall j > i: \quad & x_{e_j} = 1
\end{align}
If $e_i \in F_{GG'}$, define $x \in \{0,1\}^{E'}$ via
\begin{align}
& x_e = x_{e''} = 1 \\
& x_{C \setminus \{e,e''\}} = 0 \\
\forall j \leq i: \quad & x_{e_j} = 1 \\
\forall j > i: \quad & x_{e_j} = 0
\end{align}
Either way, it holds that $x \in S_{GG'}(e,C) \subseteq S$ and thus, $y-x \in L$. 
If $e_i \in E$, this yields
\begin{align}
0 = \sum_{1 \leq j \leq i} a_{e_j} - a_e - a_{e''} = \sum_{1 \leq j \leq i} a_{e_j}
\end{align}
by \eqref{eq:proof-cycle-facets-1}. 
If $e_i \in F_{GG'}$, we similarly obtain 
\begin{align}
0 = \sum_{i+1 \leq j \leq k} a_{e_j} - a_e - a_{e''} = \sum_{i+1 \leq j \leq k} a_{e_j}.
\end{align}
Together with \eqref{eq:proof-cycle-facets-2}, this yields $\sum_{1 \leq j \leq i} a_{e_j} = 0$ as well. 
Applying this argument repeatedly from $i=1$ to $i=m-1$, we conclude that $a_{e_1} = \dotso = a_{e_{m-1}} = 0$. 
By reversing the order of the edges in $E'_{C,v}$, it can be shown analogously that $a_{e_k} = a_{e_{k-1}} = \dotso = a_{e_{m+1}} = 0$. 
Thus, by \eqref{eq:proof-cycle-facets-2}, $a_{e'} = a_{e_m} = 0$.
\end{enumerate}

Next, consider $e' \in F_{GG'}$ and distinguish the following additional cases:
\begin{enumerate}[(i),resume]
\item \label{case:proof-cycle-facets-3} Suppose there is a $uv$-path $P'$ in $G$ that does not contain any node from $C$. Define $x \in \{0,1\}^{E'}$ via
\begin{align}
x_{e''} = \begin{cases} 0 & \text{if } e'' \in C \\ 0 & \text{if } e'' = e' \\ 0 & \text{if } e'' \in P' \text{ or } e'' \text{ is a chord of } P' \\ 1 & \text{else}. \end{cases}
\end{align}
Then $x \in S_{GG'}(e,C) \subseteq S$ and thus $y-x \in L$. This gives
\begin{align}
a_{e'} + \sum_{e'' \in P'} a_{e''} + \sum_{\substack{e'' \text{ chord}\\ \text{of } P'}} a_{e''} = 0.
\end{align}
We argue that all terms except $a_{e'}$ vanish by induction over the level function $\ell(e')$. If $\ell(e') = 1$, then $P'$ does not have any chords from $F_{GG'}$, thus $a_{e'} = 0$, because $a_{e''} = 0$ for all $e'' \in E$ as shown previously in the cases \ref{case:proof-cycle-facets-1} and \ref{case:proof-cycle-facets-2}. If $\ell(e') > 1$, then for any chord $e'' \in F_{GG'}$ of $P'$ it holds that $\ell(e'') < \ell(e')$. The induction hypothesis provides $a_{e''} = 0$ and hence we conclude $a_{e'} = 0$.
\item \label{case:proof-cycle-facets-4} Suppose $u$ is contained in $C$. Pick a shortest $uv$-path $P'$ in $G$. We argue inductively over the length of $P'$, which we denote by $d(P')$. If $d(P') = 1$, then $P'$ consists of only one edge from $E$. This situation is in fact already covered by case \ref{case:proof-cycle-facets-2}. If $d(P') > 1$, then we employ an argument similar to \ref{case:proof-cycle-facets-2} as follows.
Let $F_{C,v} := \{ \{u',v\} \in F_{GG'} \mid u' \text{ belongs to } C\} = \{f_1, \dotsc, f_k\}$ be the set of edges $f_i \in F_{GG'}$ that connect $v$ to some node in $C$. Again, assume they are ordered such that the endpoint of $f_i$ appears before the endpoint of $f_{i+1}$ on $C$ in a traversal from $e$ to itself. For the vector $x \in \{0,1\}^{E'}$ defined by
\begin{align}
x_{e''} = \begin{cases} 0 & \text{if } e'' \in C \\ 0 & \text{if } e'' \in P' \text{ or } e'' \text{ is a chord of } P' \\ 0 & \text{if } e''= u'v' \text{ where } u' \text{ belongs to } C, \\ & \hspace{55pt} v' \neq v \text{ belongs to } P' \\ 0 & \text{if } e'' \in F_{C,v} \\ 1 & \text{else}, \end{cases}
\end{align}
it holds that $x \in S_{GG'}(e,C) \subseteq S$ and thus $y-x \in L$. This yields
\begin{align}
& \sum_{e'' \in P'} a_{e''} + \sum_{\substack{e'' \text{ chord}\\ \text{of } P'}} a_{e''} \nonumber\\
+ & \!\! \sum_{\substack{e''=u'v' : \\ u' \text{ belongs to } C,\\ v' \neq v \text{ belongs to } P'}} a_{e''} + \sum_{e'' \in F_{C,v}} a_{e''} = 0
\end{align}
and thus
\begin{align}
\sum_{1\leq i \leq k} a_{f_i} = \sum_{e'' \in F_{C,v}} a_{e''} = 0, \label{eq:proof-cycle-facets-3}
\end{align}
as all other terms vanish (apply the induction hypothesis to all $u'v' \in F_{GG'}$ where $u' \text{ belongs to } C$ and $v' \neq v \text{ belongs to } P'$).
Let $m$ be the highest index such that the endpoint of $f_m$ appears before the endpoint of $P'$ on $C$. Now, for any $i$ with $1\leq i \leq m$, pick an edge $e'' \in C$ between the endpoint of $f_i$ and the endpoint of $f_{i+1}$ and before the endpoint of $P'$ on $C$. Define $x \in \{0,1\}^{E'}$ by
\begin{align}
x_{g} = \begin{cases} 0 & \text{if } g \in C \setminus \{e,e''\} \\ 0 & \text{if } g \in P' \text{ or } g \text{ is a chord of } P' \\ 0 & \text{if } g= u'v' \text{ where } \\ 
& u' \text{ appears before endpoint of } P' \text{ on } C, \\
& v' \neq v \text{ belongs to } P' \\ 0 & \text{if } g = f_j \; \forall j > i \\ 1 & \text{else}. \end{cases}
\end{align}
Then, it holds that $x \in S_{GG'}(e,C) \subseteq S$ and thus $y-x \in L$. This yields, after removing all zero terms (apply the induction hypothesis once more),
\begin{align}
\sum_{i+1 \leq j \leq k} a_{f_j} = 0.
\end{align}
Together with \eqref{eq:proof-cycle-facets-3}, we obtain
\begin{align}
\sum_{1 \leq j \leq i} a_{f_i} = 0.
\end{align}
Applying this argument repeatedly for $i=1$ to $i=m$, we conclude $a_{f_1} = \dotso = a_{f_m} = 0$. Similarly, we obtain $a_{f_k} = a_{f_{k-1}} = \dotso = a_{f_m} = 0$, by reversing the direction of traversal of $C$ and employing the same reasoning.
\item Finally, suppose neither $u$ nor $v$ belong to the cycle $C$, but every $uv$-path in $G$ shares at least one node with $C$. Let $P'$ be such a $uv$-path. Define the vector $x \in \{0,1\}^{E'}$ by
\begin{align}
x_{e''} = \begin{cases} 0 & \text{if } e'' \in C \\ 0 & \text{if } e'' = e' \\ 0 & \text{if } e'' \in P' \text{ or } e'' \text{ is a chord of } P' \\ 0 & \text{if } e''= u'v' \text{ where } u' \text{ belongs to } C, \\
& \hspace{55pt} v' \text{ belongs to } P' \\ 1 & \text{else}. \end{cases}
\end{align}
It holds that $x \in S_{GG'}(e,C) \subseteq S$ and thus $y-x \in L$. This gives
\begin{align}
a_{e'} + \sum_{e'' \in P'} a_{e''} + \sum_{\substack{e'' \text{ chord}\\ \text{of } P'}} a_{e''} + \hspace{-8pt} \sum_{\substack{e''=u'v' : \\ u' \text{ belongs to } C, \\ v' \text{ belongs to } P'}} \hspace{-8pt} a_{e''} = 0.
\end{align}
We argue inductively over the level function $\ell(e')$. If $\ell(e') = 1$, then $P'$ does not have any chords and our consideration in cases \ref{case:proof-cycle-facets-1}--\ref{case:proof-cycle-facets-4} yield that all terms except $a_{e'}$ vanish. If $\ell(e') > 1$, then we additionally employ the induction hypothesis to achieve the same result. Hence, it holds that $a_{e'} = 0$ as well.
\end{enumerate}

The proof of sufficiency in the second assertion is completely analogous (replace $C$ by $P \cup \{f\}$ and $e$ by $f$). The chosen multicuts remain valid, because $e=f$ is the only edge in the cycle that is not contained in $E$.

\begin{proposition}
\label{lemma:vwc}
For every connected graph $G = (V, E)$,
every graph $G' = (V, E')$ with $E \subseteq E'$,
every $vw \in F_{GG'}$
and every $C \in vw\cuts(G)$,
the following holds:
\begin{enumerate}[(a),nosep]
\item Every $x \in S_{GG'}(vw, C)$ defines a decomposition of $G$ into $(vw,C)$-connected components.
That is, every maximal component of the graph $(V, \{e \in E | x_e = 0\})$ is $(vw,C)$-connected.
At most one of these is properly $(vw, C)$-connected.
It exists iff $x_{vw} = 0$.
\item For every $(vw,C)$-connected component $(V^*, E^*)$ of $G$,
the $x \in \{0,1\}^{E'}$ such that
$\forall rs \in E' (x_{rs} = 0 \Leftrightarrow r \in V^* \wedge s \in V^*)$
is such that
$x \in S_{GG'}(vw, C)$.
\end{enumerate}
\end{proposition}

\paragraph{Proof of Proposition\,\ref{lemma:vwc}}
a) Let $x \in S_{GG'}(vw, C)$ arbitrary.
Let $E_0 := \{e \in E | x_e = 0\}$ and let $G_0 := (V, E_0)$.

If $x_{vw} = 1$ then $\forall e \in C: x_e = 1$, 
    by \eqref{eq:facet-discrete}.
Thus, every component of $G_0$ is improperly $(vw, C)$-connected.

If $x_{vw} = 0$ then 
\begin{align}
\exists e \in C (x_e = 0 \wedge \forall e' \in C \setminus \{e\} (x_{e'} = 1))
\label{eq:proof-lemma-vwc-1}
\end{align}
by \eqref{eq:facet-discrete}.
Let $(V^*, E^*)$ the maximal component of $G_0$ with 
\begin{align}
e \in E^*
\enspace .
\label{eq:proof-lemma-vwc-2}
\end{align}
Clearly:
\begin{align}
\forall e' \in C \setminus \{e\}:\ e' \notin E^*
\label{eq:proof-lemma-vwc-3}
\end{align}
by \eqref{eq:proof-lemma-vwc-1} and definition of $G_0$.
There does not exist a $C' \in vw\cuts(G)$ with $x_{C'} = 1$,
    because this would imply $x_{vw} = 1$, 
        by \eqref{eq:lmc-cut}.
Thus, there exists a $P \in vw\paths(G)$ with $x_{P} = 0$,
    as $G$ is connected.
Any such path $P$ has $e \in P$,
    as $P \cap C \neq \emptyset$ and $C \cap E_0 = \{e\}$ and $P \subseteq E_0$.
Thus:
\begin{align}
v \in V^* \wedge w \in V^*
\label{eq:proof-lemma-vwc-4}
\end{align}
by \eqref{eq:proof-lemma-vwc-2}.
$(V^*, E^*)$ is properly $(vw, C)$-connected, 
    by \eqref{eq:proof-lemma-vwc-2}, \eqref{eq:proof-lemma-vwc-3} and \eqref{eq:proof-lemma-vwc-4}.
Any other component of $G_0$ does not cross the cut, 
    by \eqref{eq:proof-lemma-vwc-1}, \eqref{eq:proof-lemma-vwc-2} and definition of $G_0$,
and is thus improperly $(vw, C)$-connected.

b) We have
\begin{align}
\forall st \in E:\  x_{st} = 0 \Leftrightarrow st \in E^*
\label{eq:proof-lemma-vwc-5}
\end{align}
by the following argument:
\begin{itemize}
\item If $st \in E^*$, then $s \in V^* \wedge t \in V^*$, 
    as $(V^*, E^*)$ is a graph.
Thus, $x_{st} = 0$, 
    by definition of $x$.

\item If $st \notin E^*$ then $s \notin V^* \vee t \notin V^*$,
    as $(V^*, E^*)$ is a component of $G$.
Thus, $x_{st} = 1$,
    by definition of $x$.
\end{itemize}
Consider the decomposition of $G$ into $(V^*, E^*)$ and singleton components.
$E_1 := \{e \in E|x_e = 1\}$ is the set of edges that straddle distinct components of this decomposition,
    by \eqref{eq:proof-lemma-vwc-5}.
Therefore, $E_1$ is a multicut of $G$,
    by Lemma~\ref{lemma:characteristic}.
Thus, \eqref{eq:lmc-cycle} holds,
    by Lemma~\ref{lemma:encoding}.

For any $st \in F_{GG'}$ and any $P \in st\paths(G)$, distinguish two cases:
\begin{itemize}
\item If $P \subseteq E^*$, then $s \in V^* \wedge t \in V^*$, 
    as $(V^*, E^*)$ is a graph.
Thus, $x_{st} = 0$,
    by definition of $x$.
Moreover, $x_P = 0$,
    by \eqref{eq:proof-lemma-vwc-5}.
Hence, \eqref{eq:lmc-path} evaluates to $0 = 0$.

\item Otherwise, there exists an $e \in P$ such that $e \notin E^*$.
Therefore, $x_e = 1$,
    by \eqref{eq:proof-lemma-vwc-5}.
Thus, \eqref{eq:lmc-path} holds,
    as the r.h.s.~is at least 1.
\end{itemize}

For any $st \in F_{GG'}$ and any $C' \in st\cuts(G)$, distinguish two cases:
\begin{itemize}
\item If $C' \cap E^* = \emptyset$ then $s \notin V^* \vee t \notin V^*$.
Therefore, $x_{st} = 1$,
    by definition of $x$.
Moreover, $x_{C'} = 1$,
    by \eqref{eq:proof-lemma-vwc-5}.
Thus, \eqref{eq:lmc-cut} evaluates to $0 = 0$.

\item Otherwise, there exists an $e \in C'$ such that $e \in E^*$.
Therefore, $x_e = 0$,
    by \eqref{eq:proof-lemma-vwc-5}.
Thus, \eqref{eq:lmc-cut} holds,
    as the r.h.s.~is at least 1.
\end{itemize}

\paragraph{Proof of Theorem\,\ref{theorem:facets}}
Assume that \ref{cond:cut-1} does not hold (as in Fig.~\ref{figure:violated-conditions}a).
Then, there exists an $e \in C$ such that no $(vw,C)$-connected component of $G$ contains $e$.
Thus, for all $x \in S_{GG'}(vw, C)$:
\begin{align}
x_e = 1 
\label{eq:proof-facet-1}
\end{align}
by Proposition~\ref{lemma:vwc}.
Now, $\dim \Sigma_{GG'}(vw, C) \leq |E'| - 2$,
    by \eqref{eq:facet-discrete} and \eqref{eq:proof-facet-1}.
Thus, $\Sigma_{GG'}(vw, C)$ is not a facet of $\Xi_{GG'}$,
    by Theorem~\ref{theorem:dimension}.

Assume that \ref{cond:cut-2} does not hold. Then, for any $e \in C$ there exists some number $m$ such that for all $(vw,C)$-connected components $(V^*,E^*)$ with $e \in E^*$ it holds that $\lvert F \cap F_{V^*} \rvert = m$. Thus, we can write
\begin{align}
C = \bigcup_{m=0}^{\lvert F \rvert} C(F,m),
\end{align}
where $C(F,m) := \big\{e \in C \mid \lvert F \cap F_{V^*} \rvert = m $ $\forall$ $(vw,C)$-connected $(V^*,E^*)$ with $e \in E^*\big\}$. It follows that for all $x \in S_{GG'}(vw,C)$ we have the equality
\begin{align}
\sum_{m=0}^{\lvert F \rvert} m \sum_{e \in C(F,m)} (1-x_e) = \sum_{f' \in F} (1 - x_{f'}) \label{eq:proof-facet-cond2}
\end{align}
by the following argument:
\begin{itemize}
\item If $x_e = 1$ for all $e \in C$, then $x_{f'} = 1$ for all $v'w' = f' \in F$, since $C$ is also a $v'w'$-cut. Thus, \eqref{eq:proof-facet-cond2} evaluates to $0=0$.
\item Otherwise there exists precisely one edge $e \in C$ such that $x_e = 0$. Let $m$ be such that $e \in C(F,m)$. By definition of $C(F,m)$, there are exactly $m$ edges $f' \in F$ with $x_{f'} = 0$. Thus, \eqref{eq:proof-facet-cond2} evaluates to $m=m$.
\end{itemize}

\begin{comment} % proof of old condition 3
Assume that \ref{cond:cut-3} does not hold (as in Fig.~\ref{figure:violated-conditions}e).
Then, there exist $rs = f' \in F_{GG'}(vw,C)$ and $tu = f'' \in F_{GG'}(vw,C)$ with $f' \neq f''$ such that, for every $(vw,C)$-connected component $(V^*, E^*)$ of $G$:
%
\begin{align}
r \in V^* \wedge s \in V^*
\ \Leftrightarrow\ 
t \in V^* \wedge u \in V^*
\enspace .
\label{eq:proof-facet-4}
\end{align}
%
Thus, for all $x \in S_{GG'}(vw, C)$:
%
\begin{align}
x_{f'} = x_{f''}
\label{eq:proof-facet-5}
\end{align}
%
by the following argument:
%
\begin{itemize}
\item If the decomposition of $G$ defined by $x$ (Lemma~\ref{lemma:vwc}) includes a $(vw,C)$-connected component containing both $r$ and $s$, then this component also contains $t$ and $u$, 
    by \eqref{eq:proof-facet-4}.
Thus, \eqref{eq:proof-facet-5} evaluates to $0 = 0$.
%
\item If the decomposition of $G$ defined by $x$ does not contain a $(vw,C)$-connected component containing both $r$ and $s$, then it does also not contain a $(vw,C)$-connected component containing both $t$ and $u$, 
    by \eqref{eq:proof-facet-4}.
Thus, \eqref{eq:proof-facet-5} evaluates to $1 = 1$.
\end{itemize}
%
Now, $\dim \Sigma_{GG'}(vw, C) \leq |E'| - 2$,
    by \eqref{eq:facet-discrete} and \eqref{eq:proof-facet-5}.
Thus, $\Sigma_{GG'}(vw, C)$ is not a facet of $\Xi_{GG'}$,
    by Theorem~\ref{theorem:dimension}.
\end{comment}

Assume that condition \ref{cond:cut-3} does not hold. Then there exists an $f' \in F_{GG'}(vw,C)$, a set $\emptyset \neq F \subseteq F_{GG'}(vw,C)$ and some $k \in \mathbb{N}$ such that for all $(vw,C)$ connected components $(V^*,E^*)$ and $(V^{**},E^{**})$ with $f' \in F_{V^*}$ and $f' \notin F_{V^{**}}$ it holds that
\begin{align}
\lvert F \cap F_{V^*} \rvert = k \text{ and } \lvert F \cap F_{V^{**}} \rvert = 0.
\end{align}
In other words, for all $x \in S_{GG'}(vw,C)$ it holds that $x_{f'} = 0$ iff there are exactly $k$ edges $f'' \in F$ such that $x_{f''} = 0$. Similarly, it holds that $x_{f'} = 1$ iff for all $f'' \in F$ we have $x_{f''} = 1$. Therefore, all $x \in S_{GG'}(vw,C)$ satisfy the additional equality
\begin{align}
k(1-x_{f'}) = \sum_{f'' \in F} 1 - x_{f''}.
\end{align}

\begin{figure}
\ifbool{PRECOMPILED}{
a) \imagetop{\includegraphics[scale=1]{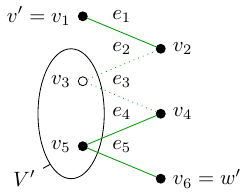}}
\hfill
b) \imagetop{\includegraphics[scale=1]{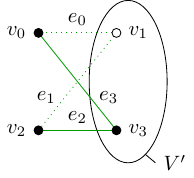}}
}{
a) \imagetop{\begin{tikzpicture}[xscale=0.6]
\draw (-1.5, 0) -- (0, 0.5); % set description
\draw[fill=white] (-0.3, 1) ellipse (0.85 and 1);
\draw[color=green] (2, 0) -- (0, 0.5) -- (2, 1);
\draw[style=cut-edge,color=green] (2, 1) -- (0, 1.5) -- (2, 2);
\draw[color=green] (2, 2) -- (0, 2.5);
\node[style=vertex,label=right:{$v_6 = w'$},fill=black] at (2, 0) {};
\node[style=vertex,label=left:$v_5$,fill=black] at (0, 0.5) {};
\node[style=vertex,label=right:$v_4$,fill=black] at (2, 1) {};
\node[style=vertex,label=left:$v_3$] at (0, 1.5) {};
\node[style=vertex,label=right:$v_2$,fill=black] at (2, 2) {};
\node[style=vertex,label=left:{$v'=v_1$},fill=black] at (0, 2.5) {};
\node at (1, 0.5) {$e_5$};
\node at (1, 1) {$e_4$};
\node at (1, 1.5) {$e_3$};
\node at (1, 2) {$e_2$};
\node at (1, 2.5) {$e_1$};
\node[fill=white] at (-1.5, 0) {$V'$};
\end{tikzpicture}}
\hfill
b) \imagetop{\begin{tikzpicture}[xscale=0.6]
\draw (2, 0) -- (3, -0.5); % set description
\draw[fill=white] (2.3, 0.75) ellipse (1 and 1.25);
\draw[color=green] (2, 0) -- (0, 0);
\draw[style=cut-edge,color=green] (0, 0) -- (2, 1.5);
\draw[style=cut-edge,color=green] (2, 1.5) -- (0, 1.5);
\draw[color=green] (0, 1.5) -- (2, 0);
\node[style=vertex,fill=black,label=right:$v_3$] at (2, 0) {};
\node[style=vertex,fill=black,label=left:$v_2$] at (0, 0) {};
\node[style=vertex,label=right:$v_1$] at (2, 1.5) {};
\node[style=vertex,fill=black,label=left:$v_0$] at (0, 1.5) {};
\node at (1, 1.7) {$e_0$};
\node at (1.8, 0.5) {$e_3$};
\node at (1, 0.2) {$e_2$};
\node at (0.2, 0.5) {$e_1$};
\node[fill=white] at (3.5, -0.5) {$V'$};
\end{tikzpicture}}
}
\caption{Depicted are the nodes (in black) and edges (in green) on a path \textbf{(a)} and on a cycle \textbf{(b)}, respectively.
Nodes in the set $V'$ are are either in $V^*$ (filled circle) or not in $V^*$ (open circle).
Consequently, pairs of consecutive edges are either cut (dotted lines) or not cut (solid lines).}
\label{figure:proof-facet-aux}
\end{figure}
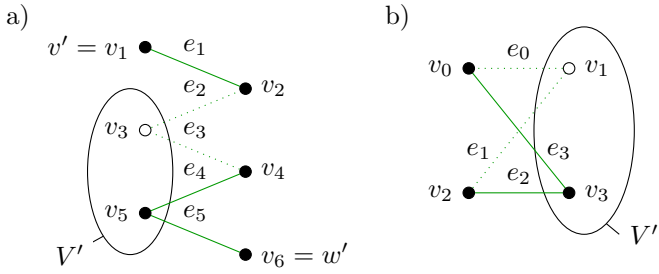

Assume that \ref{cond:cut-4} does not hold.
Then, there exist $v' \in V(v, C)$ and $w' \in V(w, C)$ and a $v'w'$-path $P = (V_P, E_P)$ in $G'(vw,C)$ such that every properly $(vw,C)$-connected component $(V^*, E^*)$ of $G$ holds:
\begin{align}
& (v' \in V^* \ \wedge\  V(w, C) \cap V_P \subseteq V^*) 
    \label{eq:proof-facet-aux2}\\
\vee \quad & (w' \in V^* \ \wedge\  V(v, C) \cap V_P \subseteq V^*) 
    \label{eq:proof-facet-aux3}
\enspace .
\end{align}
Let $v_1 < \cdots < v_{|V_P|}$ the linear order of the nodes $V_P$ and 
let $e_1 < \cdots < e_{|E_P|}$ the linear order of the edges $E_P$ 
in the $v'w'$-path $P$.
Now, for all $x \in S_{GG'}(vw, C)$:
\begin{align}
x_{vw} = \sum_{j=1}^{|E_P|} (-1)^{j+1} x_{e_j}
    \label{eq:proof-facet-6}
\end{align}
by the following argument: 
$|E_P|$ is odd,
    as the path $P$ alternates between the set $V(v, C)$ where it begins and the set $V(w, C)$ where it ends.
Thus, 
\begin{align}
\sum_{j=1}^{|E_P|} (-1)^{j+1} x_{e_j} 
    & = x_{e_1} - \sum_{j=1}^{\mathclap{(|E_P|-1)/2}} (x_{e_{2j}} - x_{e_{2j+1}})
    \enspace .
    \label{eq:proof-facet-aux4}
\end{align}
Distinguish two cases:
\begin{itemize}
\item If $x_{vw} = 1$, then $x_{E_P} = 1$, 
    by \eqref{eq:facet-discrete} and \eqref{eq:lmc-cut}.
Thus, \eqref{eq:proof-facet-6} evaluates to $1 = 1$,
    by \eqref{eq:proof-facet-aux4}.

\item If $x_{vw} = 0$, the decomposition of $G$ defined by $x$ contains precisely one properly $(vw,C)$-connected component $(V^*, E^*)$ of $G$, 
    by Proposition~\ref{lemma:vwc}.
Without loss of generality, \eqref{eq:proof-facet-aux2} holds.
    Otherwise, that is, if \eqref{eq:proof-facet-aux3} holds, exchange $v$ and $w$.

Consider the nodes $V_P$ as depicted in 
Fig.~\ref{figure:proof-facet-aux}a:
$v_1 = v' \in V^*$,
    by \eqref{eq:proof-facet-aux2}.
For every even $j$, $v_j \in V(w, C)$,
    by definition of $P$. 
Thus:
\begin{align}
\forall j \in \{1, \ldots, (|E_P| + 1)/2\}: \ 
    v_{2j} \in V^*
    \label{eq:proof-facet-aux5}
\end{align}
by \eqref{eq:proof-facet-aux2}.

Consider the edges $E_P$ as depicted in 
Fig.~\ref{figure:proof-facet-aux}a:
$e_1 = v_1 v_2 \in E^*$,
    as $v_1 \in V^*$ and $v_2 \in V^*$ and as $(V^*, E^*)$ is a component of $G$.
Thus, 
\begin{align}
x_{e_1} = 0
\label{eq:proof-facet-aux6}
\end{align}
by Proposition~\ref{lemma:vwc}.
For every $j \in \{1, \ldots, (|E_P|-1)/2\}$, distinguish two cases:
\begin{itemize}
\item If $v_{2j+1} \in V^*$, then $e_{2j} = v_{2j} v_{2j+1} \in E^*$
and $e_{2j+1} = v_{2j+1} v_{2j+2} \in E^*$,
    because $v_{2j} \in V^*$ and $v_{2j+2} \in V^*$,
        by \eqref{eq:proof-facet-aux5},
    and because $(V^*, E^*)$ is a component of $G$.
Thus:
\begin{align}
x_{e_{2j}} = 0 \ \wedge\ x_{e_{2j+1}} = 0
\enspace .
\end{align}

\item If $v_{2j+1} \notin V^*$, then $e_{2j} = v_{2j} v_{2j+1}$
and $e_{2j+1} = v_{2j+1} v_{2j+2}$ straddle distinct components of the decomposition of $G$ defined by $x$,
    because $v_{2j} \in V^*$ and $v_{2j+2} \in V^*$,
        by \eqref{eq:proof-facet-aux5}.
Thus: 
\begin{align}
x_{e_{2j}} = 1 \ \wedge\ x_{e_{2j+1}} = 1
\enspace .
\end{align}
\end{itemize}
In any case:
\begin{align}
\forall j \in \{1, \ldots, (|E_P|-1)/2\}: \ x_{e_{2j}} - x_{e_{2j+1}} = 0
\enspace .
\label{eq:proof-facet-aux7}
\end{align}
Thus, \eqref{eq:proof-facet-6} evaluates to $0 = 0$, 
    by \eqref{eq:proof-facet-aux4}, \eqref{eq:proof-facet-aux6}, \eqref{eq:proof-facet-aux7}.
\end{itemize}

Assume that \ref{cond:cut-5} does not hold.
Then, there exists a cycle $Y = (V_Y, E_Y)$ in $G'(vw, C)$
such that every properly $(vw,C)$-connected component $(V^*, E^*)$ of $G$ holds:
\begin{align}
& V_Y \cap V(v, C) \subseteq V^*
    \label{eq:proof-facet-aux8}\\
\vee \quad & V_Y \cap V(w, C) \subseteq V^*
    \label{eq:proof-facet-aux9}
\enspace .
\end{align}
Let $v_0 < \cdots < v_{|V_Y|-1}$ an order on $V_Y$ such that $v_0 \in V(v, C)$ and, for all $j \in \{0, \ldots, |E_Y|-1\}$:
\begin{align}
e_j := \{v_j, v_{j+1 \bmod |E_Y|}\} \in E_Y
\enspace .
\end{align}
Now, for all $x \in S_{GG'}(vw, C)$:
\begin{align}
0 = \sum_{j=0}^{\mathclap{|E_Y|-1}} (-1)^j x_{e_j}
    \label{eq:proof-facet-7}
\end{align}
by the following argument: 
$|E_Y|$ is even,
    as the cycle $Y$ alternates between the sets $V(v, C)$ and $V(w, C)$.
Thus, 
\begin{align}
\sum_{j=0}^{\mathclap{|E_Y|-1}} (-1)^j x_{e_j}
= \sum_{j=0}^{\mathclap{(|E_Y|-2)/2}} (x_{e_{2j}} - x_{e_{2j+1}})
\enspace .
\label{eq:proof-facet-aux10}
\end{align}
Distinguish two cases:
\begin{itemize}
\item If $x_{vw} = 1$, then $x_{E_Y} = 1$, 
    by \eqref{eq:facet-discrete} and \eqref{eq:lmc-cut}.
Thus, \eqref{eq:proof-facet-7} evaluates to $0 = 0$,
    by \eqref{eq:proof-facet-aux10}.

\item If $x_{vw} = 0$, the decomposition of $G$ defined by $x$ contains precisely one properly $(vw,C)$-connected component $(V^*, E^*)$ of $G$, 
    by Proposition~\ref{lemma:vwc}.
Without loss of generality, \eqref{eq:proof-facet-aux8} holds.
    Otherwise, that is, if \eqref{eq:proof-facet-aux9} holds, exchange $v$ and $w$.

Consider the nodes $V_Y$ as depicted in 
Fig.~\ref{figure:proof-facet-aux}b:
For every even $j$, $v_j \in V(v, C)$,
    by definition of $Y$ and the order. 
Thus:
\begin{align}
\forall j \in \{0, \ldots, (|E_Y|-2)/2\}: \ 
    v_{2j} \in V^*
    \label{eq:proof-facet-aux11}
\end{align}
by \eqref{eq:proof-facet-aux8}.

Consider the edges $E_Y$ as depicted in 
Fig.~\ref{figure:proof-facet-aux}b:
For every $j \in \{0, \ldots, (|E_Y|-2)/2\}$, distinguish two cases:
\begin{itemize}
\item If $v_{2j+1} \in V^*$, then 
$e_{2j} = v_{2j} v_{2j+1} \in E^*$ and 
$e_{2j+1} = v_{2j+1} v_{2j+2 \bmod |E_Y|} \in E^*$,
    because $v_{2j} \in V^*$ and $v_{2j+2 \bmod |E_Y|} \in V^*$,
        by \eqref{eq:proof-facet-aux11},
    and because $(V^*, E^*)$ is a component of $G$.
Thus:
\begin{align}
x_{e_{2j}} = 0 \ \wedge\ x_{e_{2j+1}} = 0
\enspace .
\end{align}

\item If $v_{2j+1} \notin V^*$, then
$e_{2j} = v_{2j} v_{2j+1}$ and 
$e_{2j+1} = v_{2j+1} v_{2j+2 \bmod |E_Y|}$
straddle distinct components of the decomposition of $G$ defined by $x$,
    because $v_{2j} \in V^*$ and $v_{2j+2 \bmod |E_Y|} \in V^*$,
        by \eqref{eq:proof-facet-aux11}.
Thus: 
\begin{align}
x_{e_{2j}} = 1 \ \wedge\ x_{e_{2j+1}} = 1
\enspace .
\end{align}
\end{itemize}
In any case:
\begin{align}
\forall j \in \{0, \ldots, (|E_Y|-2)/2\}: \ x_{e_{2j}} - x_{e_{2j+1}} = 0
\enspace .
\label{eq:proof-facet-aux12}
\end{align}
Thus, \eqref{eq:proof-facet-7} evaluates to $0 = 0$, 
    by \eqref{eq:proof-facet-aux10} and \eqref{eq:proof-facet-aux12}.
\end{itemize}

\begin{figure*}
\ifbool{PRECOMPILED}{
a) \imagetop{\includegraphics[scale=1]{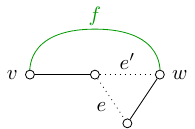}}
\hfill
b) \imagetop{\includegraphics[scale=1]{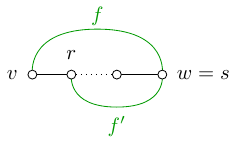}}
\hfill
c) \imagetop{\includegraphics[scale=1]{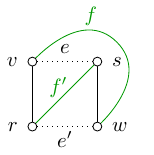}}
d) \imagetop{\includegraphics[scale=1]{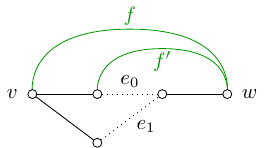}}
\\
e) \imagetop{\includegraphics[scale=1]{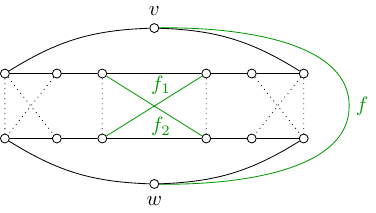}}
\hfill
f) \imagetop{\includegraphics[scale=1]{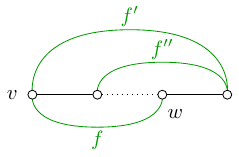}}
\hfill
g) \imagetop{\includegraphics[scale=1]{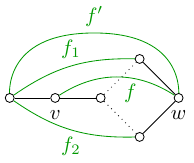}}
\\
h) \imagetop{\includegraphics[scale=1]{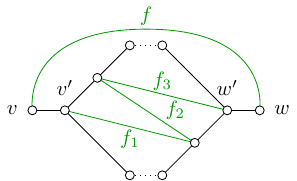}}
\hfill
i) \imagetop{\includegraphics[scale=1]{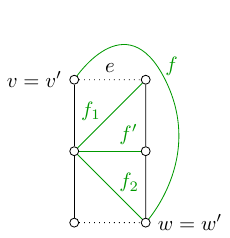}}
\hfill
j) \imagetop{\includegraphics[scale=1]{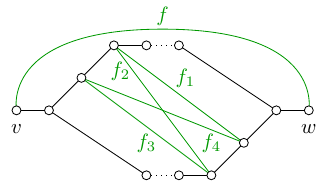}}
\\
k) \imagetop{\includegraphics[scale=1]{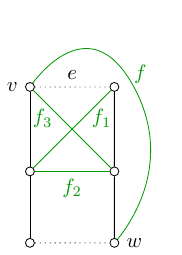}}
}{
\makebox(0,0){a)}
\input{figures/cuts-examples-a.tex}
\hfill
\makebox(0,0){b)}
\input{figures/cuts-examples-b.tex}
\hfill
\makebox(0,0){c)}
\input{figures/cuts-examples-c.tex}
\makebox(0,0){d)}
\input{figures/cuts-examples-d.tex}
\\
\makebox(0,0){e)}
\begin{subtikzpicture}{e)}
\draw[green] plot [smooth, tension=2] coordinates {(0, -0.3) (3,-1.5) (0,-2.7)};
\node at (3.2,-1.5) {$\color{green}f$};
\node[style=vertex,label=above:{$v$}] (v) at (0,-0.3) {};
\node[style=vertex] (1) at (2.3,-1) {};
\node[style=vertex] (2) at (1.5,-1) {};
\node[style=vertex] (3) at (0.8,-1) {};
\node[style=vertex] (4) at (-0.8,-1) {};
\node[style=vertex] (5) at (-1.5,-1) {};
\node[style=vertex] (6) at (-2.3,-1) {};
\node[style=vertex] (7) at (2.3,-2) {};
\node[style=vertex] (8) at (1.5,-2) {};
\node[style=vertex] (9) at (0.8,-2) {};
\node[style=vertex] (10) at (-0.8,-2) {};
\node[style=vertex] (11) at (-1.5,-2) {};
\node[style=vertex] (12) at (-2.3,-2) {};
\node[style=vertex,label=below:{$w$}] (w) at (0,-2.7) {};
\draw[-] (v) edge [bend left=15] (1);
\draw[-] (v) edge [bend right=15] (6);
\draw[-] (w) edge [bend right=15] (7);
\draw[-] (w) edge [bend left=15] (12);
\draw[-] (1) edge (2);
\draw[-] (2) edge (3);
\draw[-] (3) edge (4);
\draw[-] (4) edge (5);
\draw[-] (5) edge (6);
\draw[-] (7) edge (8);
\draw[-] (8) edge (9);
\draw[-] (9) edge (10);
\draw[-] (10) edge (11);
\draw[-] (11) edge (12);
\draw[style=cut-edge] (1) edge (7);
\draw[style=cut-edge] (1) edge (8);
\draw[style=cut-edge] (2) edge (7);
\draw[style=cut-edge] (3) edge (9);
\draw[style=cut-edge] (4) edge (10);
\draw[style=cut-edge] (5) edge (12);
\draw[style=cut-edge] (6) edge (11);
\draw[style=cut-edge] (6) edge (12);
\draw[-,green] (3) edge node[above=10,right=-5] {$f_1$} (10);
\draw[-,green] (4) edge node[below=10,right=-5] {$f_2$} (9);
\end{subtikzpicture}
\hfill
\makebox(0,0){f)}
\input{figures/cuts-examples-f.tex}
\hfill
\makebox(0,0){g)}
\begin{subtikzpicture}{g)}
\draw[green] plot [smooth, tension=2] coordinates {(0, 0) (1.3, 1) (2.6, 0)};
\node at (1.3, 1.25) {$\color{green}f'$};
\node[style=vertex] (1) at (0, 0) {};
\node[style=vertex,label=below:$v$] (2) at (0.7, 0) {};
\node[style=vertex] (3) at (1.4, 0) {};
\node[style=vertex] (4) at (2, 0.6) {};
\node[style=vertex] (5) at (2, -0.6) {};
\node[style=vertex,label=below:$w$] (6) at (2.6, 0) {};
\draw (1) -- (2);
\draw (2) -- (3);
\draw[style=cut-edge] (3) -- (4);
\draw[style=cut-edge] (3) -- (5);
\draw (4) -- (6);
\draw (5) -- (6);
\draw[-,green] (1) edge [bend left=18] node[above] {$f_1$} (4);
\draw[-,green] (1) edge [bend right=18] node[below] {$f_2$} (5);
\draw[-,green] (2) edge [bend left=33] node[below=8,right=0] {$f$} (6);
\end{subtikzpicture}
\\
\makebox(0,0){h)}
\input{figures/cuts-examples-h.tex}
\hfill
\makebox(0,0){i)}
\begin{subtikzpicture}{i)}
\draw[green] plot [smooth, tension=2] coordinates {(0, 1.1) (1.4, 1.1) (1.1, -1.1)};
\node at (1.5,1.3) {$\color{green}f$};
\node[style=vertex,label=left:{$v=v'$}] (v) at (0,1.1) {};
\node[style=vertex] (1) at (0,0) {};
\node[style=vertex] (2) at (0,-1.1) {};
\node[style=vertex] (3) at (1.1,1.1) {};
\node[style=vertex] (4) at (1.1,0) {};
\node[style=vertex,label=right:{$w=w'$}] (w) at (1.1,-1.1) {};
\draw[-,green] (1) edge node[above=2,left=1] {$f_1$} (3);
\draw[-,green] (1) edge node[above=2,right=1] {$f_2$} (w);
\draw[-,green] (1) edge node[above=8,right=1] {$f'$} (4);
\draw[-] (v) edge (1);
\draw[-] (1) edge (2);
\draw[-] (3) edge (4);
\draw[-] (4) edge (w);
\draw[style=cut-edge] (v) edge node[above] {$e$} (3);
\draw[style=cut-edge] (2) edge (w);
\end{subtikzpicture}
\hfill
\makebox(0,0){j)}
\input{figures/cuts-examples-j.tex}
\\
\makebox(0,0){k)}
\begin{subtikzpicture}{k)}
\draw[green] plot [smooth, tension=2] coordinates {(0, 1.3) (1.6, 1.3) (1.3, -1.1)};
\node at (1.7,1.5) {$\color{green}f$};
\node[style=vertex,label=left:{$v$}] (v) at (0,1.3) {};
\node[style=vertex] (1) at (0,0) {};
\node[style=vertex] (2) at (0,-1.1) {};
\node[style=vertex] (3) at (1.3,1.3) {};
\node[style=vertex] (4) at (1.3,0) {};
\node[style=vertex,label=right:{$w$}] (w) at (1.3,-1.1) {};
\draw[-,green] (v) edge node[above=5,left=6] {$f_3$} (4);
\draw[-,green] (1) edge node[above=5,right=6] {$f_1$} (3);
\draw[-,green] (1) edge node[below] {$f_2$} (4);
\draw[-] (v) edge (1);
\draw[-] (1) edge (2);
\draw[-] (3) edge (4);
\draw[-] (4) edge (w);
\draw[style=cut-edge] (v) edge node[above] {$e$} (3);
\draw[style=cut-edge] (2) edge (w);
\end{subtikzpicture}
}
\caption{Depicted above are graphs $G = (V, E)$ (in black)
and $G' = (V, E')$ with $E \subseteq E'$ ($E'$ in green),
distinct nodes $v, w \in V$ and a $vw$-cut $C$ of $G$ (as dotted lines).
In any of the above examples, one condition of 
Theorem~\ref{theorem:facets}
is violated and thus, 
$\Sigma_{GG'}(vw, C)$ is not a facet of the lifted multicut polytope $\Xi_{GG'}$.
\textbf{a)} Condition \ref{cond:cut-1} is violated for $e$.
\textbf{b)} Condition \ref{cond:cut-2} is violated as $r$ and $s$ are connected in any $(vw, C)$-connected component.
\textbf{c)} Condition \ref{cond:cut-2} is violated as $r$ and $s$ are not connected in any $(vw, C)$-connected component.
\textbf{d)} Condition \ref{cond:cut-2} is violated. Specifically, $C(\{f'\},1) = \{e_0\}$ and $C(\{f'\},0) = \{e_1\}$ in the proof of
Theorem~\ref{theorem:facets}.
\textbf{e)} Condition \ref{cond:cut-2} is violated for $F= \{f_1,f_2\}$.
\textbf{f)} Condition \ref{cond:cut-3} is violated.
\textbf{g)} Condition \ref{cond:cut-3} is violated for $F = \{f_1,f_2\}$ and $k=1$.
\textbf{h)} Condition \ref{cond:cut-4} is violated for the $v'w'$-path $f_1 f_2 f_3$.
\textbf{i)} Condition \ref{cond:cut-4} is violated for the $v'w'$-path $e f_1 f_2$.
\textbf{j)} Condition \ref{cond:cut-5} is violated for the cycle $f_1 f_2 f_3 f_4$.
\textbf{k)} Condition \ref{cond:cut-5} is violated for the cycle $e f_1 f_2 f_3$.
}
\label{figure:violated-conditions}
\end{figure*}

\end{document}